\documentclass[]{JHEP3}
\pdfoutput=1
\usepackage{array}
\usepackage{epsfig}
\usepackage{amssymb}
\usepackage{graphics}
\usepackage{slashed}
\usepackage{amsmath}

\newcommand{\bal}{\begin{align}}

\def \kstar{{K^*}}
\def \eff{{\text{eff}}}
\allowdisplaybreaks

\title{New strategies for New Physics search in \boldmath $B\to K^*\nu\bar\nu$, $B\to K\nu\bar\nu$ and $B\to X_s\nu\bar\nu$ decays}

\author{Wolfgang~Altmannshofer$^{a}$,
Andrzej~J.~Buras$^{ab}$,
David~M.~Straub$^{a}$ and
Michael~Wick$^{a}$
\\
$^a$ Physik-Department, Technische Universit\"at M\"unchen,\\ James-Franck-Str., 85748 Garching, Germany\\
$^b$ TUM Institute for Advanced Study, Technische Universit\"at M\"unchen,\\ Arcisstr. 21, 80333 M\"unchen, Germany
\\
E-Mail: 
\email{wolfgang.altmannshofer@ph.tum.de},
\email{andrzej.buras@ph.tum.de},
\email{david.straub@ph.tum.de},
\email{michael.wick@ph.tum.de}
}

\preprint{TUM-HEP-709/09}

\abstract{
The rare decay $B\to K^*\nu \bar \nu$ is regarded as one of the important channels in $B$ physics as it allows a transparent study of $Z$ penguin and other electroweak penguin effects in New Physics (NP) scenarios in the absence of dipole operator contributions and Higgs (scalar) penguin contributions that are often more important than $Z$ contributions in $B\to K^*\ell^+\ell^-$ and $B_s\to \ell^+\ell^-$ decays. We present a new analysis of $B\to K^*\nu\bar\nu$ with improved form factors and of the decays $B\to K\nu\bar\nu$ and $B\to X_s\nu\bar\nu$ in the SM and in a number of NP scenarios like the general MSSM, general scenarios with modified $Z$/$Z'$ penguins and in a singlet scalar extension of the SM. We also summarize the results in the Littlest Higgs model with T-parity and a Randall-Sundrum (RS) model with custodial protection of left-handed $Z d_i \bar d_j$ couplings.
Our SM prediction $\text{BR}(B \to K^* \nu\bar\nu)=( 6.8^{+1.0}_{-1.1} ) \times 10^{-6}$ turns out to be significantly lower than the ones present in the literature.
Our improved calculation $\text{BR}(B\to X_s\nu\bar\nu)=(2.7\pm0.2) \times 10^{-5}$ in the SM avoids the normalization to the $\text{BR}(B\to X_ce\bar\nu_e)$ and, with less than 10\% total uncertainty, is the most accurate to date.
The results for the SM and NP scenarios can be transparently summarized in a $(\epsilon,\eta)$ plane analogous to the known $(\bar\varrho,\bar\eta)$ plane with a non-vanishing $\eta$ signalling this time not CP violation but the presence of new right-handed down-quark flavour violating couplings which can be ideally probed by the decays in question. Measuring the three branching ratios and one additional polarization observable in $B\to K^*\nu \bar \nu$ allows to overconstrain the resulting point in the $(\epsilon,\eta)$ plane with $(\epsilon,\eta)=(1,0)$ corresponding to the SM. We point out that the correlations of these three channels with the rare decays $K^+\to\pi^+\nu\bar\nu$, $K_L\to\pi^0\nu\bar\nu$, $B\to X_s \ell^+\ell^-$ and $B_s\to \mu^+\mu^-$ offer powerful tests of New Physics with new right-handed couplings and non-MFV interactions.}


\begin{document}

\section{Introduction}

Rare $K$ and $B$ decays with a $\nu\bar\nu$ pair in the final state belong to the theoretically cleanest decays in the field of flavour changing neutral current (FCNC) processes. Indeed, the presence of $\nu\bar\nu$ in the final states eliminates in the case of inclusive decays non-perturbative contributions related to low energy QCD dynamics and photon exchanges and in the case of exclusive decays allows to encode efficiently such contributions in the hadronic matrix elements of quark currents. In the case of the rare decays $K^+\to\pi^+\nu\bar\nu$ and $K_L\to\pi^0\nu\bar\nu$ these matrix elements can be extracted from the data on the leading semi-leptonic $K^+$ and $K_L$ decays using isospin symmetry. On the other hand, the study of the exclusive decays $B\to K^*\nu\bar\nu$ and $B\to K\nu\bar\nu$ requires the evaluation of the relevant form factors by means of non-perturbative methods.

Over the last twenty years, extensive analyses of the decays
$K^+\to\pi^+\nu\bar\nu$ and $K_L\to\pi^0\nu\bar\nu$ have been performed
in the literature. Most recent reviews can be found in \cite{Isidori:2007zs,Buras:2004uu,Smith:2006qg}. Moreover,
seven events of $K^+\to\pi^+\nu\bar\nu$ have been reported \cite{Artamonov:2008qb}. While a number
of analyses of $B\to K^*\nu\bar\nu$, $B\to K\nu\bar\nu$ and 
$B\to X_{s}\nu\bar\nu$ appeared already in the literature, we think that
the power of these decays in testing the short distance physics 
related in particular to $Z$ penguin diagrams has not been fully appreciated
yet, possibly due to great challenges in measuring their branching ratios.
With the advent of Super-B facilities \cite{Bona:2007qt}, the prospects of measuring these
branching ratios in the next decade are not fully unrealistic and it seems
appropriate to have a closer look at these decays in order to motivate
further experimental efforts to measure their branching ratios and related
observables.

While the decay $B\to K^*\nu\bar\nu$ is theoretically not as clean as 
$K^+\to\pi^+\nu\bar\nu$ and $K_L\to\pi^0\nu\bar\nu$ because of the presence
of form factors that have to be calculated by non-perturbative methods, it
should be emphasized that the existence of angular observables in 
$B\to K^*\nu\bar\nu$ allows a deeper insight into the issue of right-handed
currents than it is possible in the two rare $K$ decays in question. Indeed
the latter decays are only sensitive to the sum of the 
Wilson coefficients of left-handed and right-handed couplings, whereas
$B\to K^*\nu\bar\nu$ is also sensitive to their difference.

In a recent paper \cite{Altmannshofer:2008dz}, we have presented a detailed study of angular observables in the rare decay $B\to K^*(\to K\pi)\mu^+\mu^-$, demonstrating its outstanding virtues in testing the Standard Model (SM) and its extentions. Other recent analyses of $B\to K^*(\to K\pi)\mu^+\mu^-$ can be found in \cite{Egede:2008uy,Bobeth:2008ij}. The goal of the present paper is to extend our study to $B\to K^*(\to K\pi)\nu\bar\nu$, making use of the relevant form factors discussed in detail in our analysis \cite{Altmannshofer:2008dz}, where various extensions of the SM have already been described.

In the SM and in models with minimal flavour violation (MFV) there is a striking correlation
 between the branching ratios for $K_L\to\pi^0\nu\bar\nu$ and $B\to X_s\nu\bar\nu$ as the same
 one-loop function
$X(x_t)$ governs the two processes in question \cite{Buras:2001af}. 
This relation is generally 
modified in models with non-MFV interactions. As we will see below 
there are also correlations between $K_L\to\pi^0\nu\bar\nu$, $K^+\to\pi^+ \nu\bar\nu$  
and $B\to K^*(\to K\pi)\nu\bar\nu$ that
are useful for the study of various NP scenarios.

Our paper is organized as follows. In section \ref{sec:obs} we recall the effective Hamiltonian for $b\to s \nu\bar\nu$ transitions and define the observables that can in principle be measured in $B\to K^*(\to K\pi)\nu\bar\nu$, $B\to K\nu\bar\nu$ and $B\to X_s\nu\bar\nu$. In section \ref{sec:numerics} we present a numerical analysis of these decays, first within the SM and then beyond, both model-independently and within concrete extensions of the SM. We summarize our results in section \ref{sec:sum}, stressing the novel features of our analysis.


\section{Exclusive and inclusive {\boldmath $b\to s\nu\bar\nu$} decays}\label{sec:obs}

In this section we summarize the effective Hamiltonian for $b\to s\nu\bar\nu$ transitions and collect all $B$ decays probing this quark level transition. Our focus is on the decay $B\to K^*\nu\bar\nu$ which, due to its additional polarization observable, offers a richer source of information than the two other decays $B\to K\nu\bar\nu$ and $B\to X_s\nu\bar\nu$. Combining all decays we end up with {\it four} observables which are functions of the invariant mass of the neutrino-antineutrino pair.

\subsection{Effective Hamiltonian}

The effective Hamiltonian for $b\to s \nu\bar\nu$ transitions is generally given by
\begin{equation} \label{eq:Heff}
{\cal H}_{\eff} = - \frac{4\,G_F}{\sqrt{2}}V_{tb}V_{ts}^*\left(C^\nu_L \mathcal O^\nu_L +C^\nu_R \mathcal O^\nu_R  \right) ~+~ {\rm h.c.} ~,
\end{equation}
with the operators
\begin{align}
\mathcal{O}^\nu_{L} &=\frac{e^2}{16\pi^2}
(\bar{s}  \gamma_{\mu} P_L b)(  \bar{\nu} \gamma^\mu(1- \gamma_5) \nu)~,&
\mathcal{O}^\nu_{R} &=\frac{e^2}{16\pi^2}(\bar{s}  \gamma_{\mu} P_R b)(  \bar{\nu} \gamma^\mu(1- \gamma_5) \nu)~.
\end{align}
In the SM, $C^\nu_R$ is negligible while $C^\nu_L=-X(x_t)/\sin^2\theta_w$, where $x_t=m_t^2/m_W^2$ and the function $X(x_t)$ can be found in ref.~\cite{Misiak:1999yg,Buchalla:1998ba} at the next-to-leading order in QCD.

Taking into account the latest top mass measurement from the Tevatron~\cite{:2008vn}, we obtain
\begin{equation}
(C^\nu_L)^\text{SM} = - 6.38 \pm 0.06~,
\label{eq:CLSM}
\end{equation}
where the error is dominated by the top mass uncertainty. The corresponding operator is not renormalized by QCD, so the only renormalization scale dependence enters $X(x_t)$ through the running top quark mass, which is however largely cancelled through NLO QCD corrections. The residual scale dependence is taken into account in the error in eq.~(\ref{eq:CLSM}).

\subsection[$B\to K^*\nu\bar\nu$]{{\boldmath$B\to K^*\nu\bar\nu$}}\label{sec:BKsnunu}

The decay $B\to K^*\nu\bar\nu$ has the virtue that the angular distribution of the $K^*$ decay products allows to extract information about the polarization of the $K^*$, just like in $B\to K^*\mu^+\mu^-$ decays.
Since the neutrinos escape the detector unmeasured, the experimental information that can be obtained from the process $B\to K^* (\to K\pi)\nu\bar\nu$ with an on-shell $K^*$ is completely described by the double differential decay distribution in terms of the two kinematical variables $s_B=q^2/m_B^2$, where $q^2$ is the invariant mass of the neutrino-antineutrino pair, and $\theta$, the angle between the $K^*$ flight direction in the $B$ rest frame and the $K$ flight direction in the $K\pi$ rest frame. The normalized invariant mass $s_B$ ranges from 0 to the kinematical endpoint $(1-\widetilde{m}_{K^*})^2\approx0.69$, where here and in the following we use $\widetilde{m}_i=m_i/m_B$, while $\theta$ ranges from 0 to $\pi$.

The spectrum can be expressed in terms of $B\to K^*$ transversity amplitudes $A_{\perp,\parallel,0}$, which are given in terms of form factors and Wilson coefficients as

\begin{align}
A_{\perp}(s_B)  &=  2 N \sqrt{2}  \lambda^{1/2}(1,\widetilde{m}_{K^*}^2,s_B) (C^\nu_L+C^\nu_R) \frac{ V(s_B) }{(1 + \widetilde{m}_\kstar)}~, \\
A_{\parallel}(s_B)  & = - 2 N \sqrt{2}(1 + \widetilde{m}_\kstar) (C^\nu_L-C^\nu_R) A_1(s_B)~, \\
A_{0}(s_B)  &=  - \frac{N (C^\nu_L-C^\nu_R)}{\widetilde{m}_\kstar  \sqrt{s_B }} 
\bigg[(1 - \widetilde{m}_\kstar^2 - s_B) ( 1 + \widetilde{m}_\kstar) A_1(s_B) 
 -\lambda(1,\widetilde{m}_{K^*}^2,s_B) \frac{A_2(s_B)}{1 + \widetilde{m}_\kstar}
\bigg]~,
\end{align}
where
\begin{equation}
N= V_{tb}^{\vphantom{*}}V_{ts}^* \left[\frac{G_F^2 \alpha^2 m_B^3}{3\cdot 2^{10}\pi^5 }
 s_B  \lambda^{1/2}(1,\widetilde{m}_{K^*}^2,s_B)
\right]^{1/2}
\end{equation}
and $\lambda(a,b,c)= a^2  +b^2 + c^2 - 2 (a b+ b c + a c)$.

The analysis in our paper is done with $B \to K^*$ form factors $V(q^2)$, $A_1(q^2)$ and $A_2(q^2)$, which are based on the low-$q^2$ form factors given in \cite{Altmannshofer:2008dz}, which are calculated from QCD sum rules on the light cone.
For the high $q^2$ region, where the light-cone expansion breaks down, we adopt an extrapolation following the steps of \cite{Ball:2004rg}. There the low-$q^2$ form factors, which are obtained from light-cone sum rules as well, are fitted to parametrizations accounting for resonances in the form factors. To estimate the dependence of our analysis on the form factors, we will confront in section \ref{sec:SM} some of our results with the results using two older sets of form factors given in the literature.

Defining the invariant mass spectrum with a longitudinally and transversely polarized $K^*$, respectively, as
\begin{equation}
\frac{d\Gamma_L}{ds_B} = 3 m_B^2|A_0|^2 ~,
\qquad
\frac{d\Gamma_T}{ds_B} = 3  m_B^2\left( |A_\perp|^2 + |A_\parallel|^2 \right) ~,
\end{equation}
where the factor of 3 stems from the sum over neutrino flavours\footnote{Here we assume that the Wilson coefficients do not depend on the neutrino flavour, which is an excellent approximation in all the models we consider in sec.~\ref{sec:numerics}.}, the double differential spectrum can be written as
\begin{equation}
\frac{d^2\Gamma}{ds_B d\!\cos\!\theta} =  \frac{3}{4} \frac{d\Gamma_T}{ds_B} \sin^2\theta + \frac{3}{2} \frac{d\Gamma_L}{ds_B} \cos^2\theta~.
\label{eq:doublediff}
\end{equation}
Thus, $d\Gamma_L/ds_B$ and $d\Gamma_T/ds_B$ can be extracted by an angular analysis of the $K^*$ decay products.

Instead of these two observables, one can choose the following two independent observables accessible from the double differential decay distribution: the dineutrino mass distribution $d\Gamma/ds_B$, where
\begin{equation}
\frac{d\Gamma}{ds_B } = \int_{-1}^{1} d\!\cos\!\theta \, \frac{d^2\Gamma}{ds_B d\!\cos\!\theta} = \frac{d\Gamma_L}{ds_B } + \frac{d\Gamma_T}{ds_B} =  3 m_B^2 \left( |A_\perp|^2 + |A_\parallel|^2 + |A_0|^2 \right)~,
\end{equation}
and either of the $K^*$ longitudinal and transverse polarization fractions $F_{L,T}$ also used in studies of $B\to K^*\ell^+\ell^-$ decays and defined as
\begin{equation}
F_{L,T} = \frac{d\Gamma_{L,T}/ds_B}{d\Gamma/ds_B}~,~~~F_L = 1 - F_T~.
\end{equation}
The advantage of this choice of observables is twofold. First, the normalization of $F_{L,T}$ on the total dineutrino spectrum strongly reduces the hadronic uncertainties associated with the form factors as well as parametric uncertainties associated with CKM elements. Second, in the absence of right-handed currents ($C^\nu_R=0$), the dependence on the remaining Wilson coefficient $C^\nu_L$ drops out in $F_{L,T}$, making it a perfect observable to probe such right-handed currents.

In section~\ref{sec:indep}, we will also consider the $s_B$-integrated form of $F_{L,T}$, which we define as
\begin{equation}
\langle F_{L,T} \rangle = \frac{\Gamma_{L,T}}{\Gamma},
\qquad\text{where}\qquad
\Gamma_{(L,T)} = \int_0^{1-\widetilde{m}_{K^*}^2} ds_B \frac{d\Gamma_{(L,T)}}{ds_B} ~.
\end{equation}

As a final note, we emphasize that the transverse asymmetry
\begin{equation}
A_T = \frac{-2\text{Re}(A_\perp A_\parallel^*)}{|A_\perp|^2 + |A_\parallel|
^2}
\end{equation}
which was studied in \cite{Melikhov:1998ug} cannot be extracted from a measurement of the angular distribution of $B\to K^* (\to K\pi)\nu\bar\nu$ \cite{Kim:1999waa} as this would require a measurement of the neutrino polarization, which is clearly impossible. This fact was discussed in ref.~\cite{Egede:2008uy} in the context of $B\to K^* (\to K\pi)\ell^+\ell^-$, where the corresponding asymmetry is denoted $A_T^{(1)}$.

\subsection[$B\to K\nu\bar\nu$]{\boldmath $B\to K\nu\bar\nu$}

The dineutrino invariant mass distribution for the exclusive decay $B\to K 
\nu\bar\nu$ can be written as \cite{Colangelo:1996ay}
\begin{equation}
  \frac{d \Gamma (B \to K \nu\bar\nu)}{ds_B }  =  
  \frac{G_F^2  \alpha^2 }{256 \pi^5 } 
      \left| V_{ts}^*  V_{tb} \right|^2 m_B^5\lambda^{3/2}(s_B,\widetilde{m}_K^2,1)
\left[f^K_+(s_B)\right]^2 \left|C^\nu_L +C^\nu_R \right|^2~.
\end{equation}

We use the $B\to K$ form factor $f^K_+$ given in \cite{Ball:2004ye}, which is valid in the full physical regime $0 \le s_B  \le (1  -\widetilde{m}_K )^2  \approx 0.82$. As argued by the authors of \cite{Ball:2004ye}, we assume that the maximum uncertainty is at $s_B=0$ and, to be conservative, we adopt this uncertainty for the full $s_B$ range.

\subsection[$B\to X_s\nu\bar\nu$]{\boldmath $B\to X_s\nu\bar\nu$}\label{sec:BXsnunu}
The decay $B\to X_s\nu\bar\nu$ offers the theoretically cleanest constraint on the Wilson coefficients $C^{\nu}_L$ and $C^{\nu}_R$ as it does not involve any form factors. Its dineutrino invariant mass distribution is sensitive to yet another combination of $C_L^\nu$ and $C_R^\nu$,
\begin{multline}\label{eq:BXsnn}
\frac{d\Gamma(B \to  X_s\nu\bar\nu)}{ds_b} =
m_b^5\frac{\alpha^2 G_F^2}{128 \pi^5}
|V_{ts}^*V_{tb}|^2 \kappa(0) (|C^\nu_{L}|^2 + |C^\nu_{R}|^2)\\
\times \sqrt{\lambda(1,\hat m_s^2,s_b)}\left[3 s_b (1 + \hat m_s^2 - s_b - 4 \hat m_s \frac{\text{Re}\left(C^\nu_L C_R^{\nu *}\right)}{|C^\nu_L|^2 + |C^\nu_R|^2}) +\lambda(1,\hat m_s^2,s_b)\right]~,
\end{multline}
where we have defined $\hat m_i = m_i/m_b$ and $\kappa(0)=0.83$ represents the QCD correction to the $b\to s\nu\bar\nu$ matrix element \cite{Grossman:1995gt,Buchalla:1995vs,Bobeth:2001jm}.

In previous analyses of $B \to X_s \nu\bar\nu$, similar to the practice in the calculation of BR$(B \to X_s \gamma)$ \cite{Chetyrkin:1996vx}, the common approach to reduce the theoretical uncertainties was to normalize eq.~(\ref{eq:BXsnn}) to the inclusive semileptonic decay rate $\Gamma(B\to X_c e \bar \nu_e)$ to avoid the overall factor of $m_b^5$. However, in this approach an additional uncertainty is introduced through the dependence of the semileptonic phase space factor on the charm quark mass. (See e.g. \cite{Gambino:2001ew,Gambino:2008fj} on how to address this problem in the case of the $B \to X_s \gamma$ decay.)

In the case of $B \to X_s \nu\bar\nu$, we can go even further and adopt a novel approach\footnote{%
We are indebted to Miko{\l}aj Misiak for suggesting this strategy.} 
by refraining totally from this normalization and directly using eq.~(\ref{eq:BXsnn}) in combination with the $b$ quark mass in the 1S scheme, which is known to a precision of about 1\% \cite{Hoang:1998ng,Hoang:1998hm,Hoang:2000fm,Bauer:2004ve}. For the branching ratio, which is obtained by integrating eq. (\ref{eq:BXsnn}) over the kinematically allowed region $0\le s_b\le(1-\hat m_s)^2\approx0.96$, taking into account the additional $O(\Lambda^2/m_b^2)$ corrections \cite{Falk:1995me,Grossman:1995gt} with the HQET parameters taken from \cite{Bauer:2004ve}, we thus obtain an estimated uncertainty of less than 10\%. This constitutes a considerable improvement compared to the conventional approach.

Our choices of errors will be presented in more detail in section \ref{sec:SM}.

\section{Numerical analysis}\label{sec:numerics}

In this section, we discuss our predictions for the four $b\to s\nu\bar\nu$ observables defined in the previous section, i.e. three branching ratios and the angular observable $F_L(B\to K^*\nu\bar\nu)$. The input values for the parameters used in the numerical analysis are collected in table~\ref{tab:par}.  For the branching ratio predictions, we use the $B^\pm$ lifetime $\tau_{B^+}$ for the $B\to K$ decays, the $B^0$ lifetime $\tau_{B^0}$ for the $B\to K^*$ decays and their average $\tau_B = (\tau_{B^+}+\tau_{B^0})/2$ for the inclusive decays.

After updating the SM predictions in section~\ref{sec:SM}, we discuss NP effects on the Wilson coefficients in a model-independent manner in section~\ref{sec:indep} and under the assumption of $Z$ or $Z'$ penguin dominance in section~\ref{sec:penguins}, briefly comment on the Littlest Higgs model with T-parity and RS model with custodial protection of left-handed $Z$ couplings in sections~\ref{NPLHT} and \ref{sec:WED} and discuss in detail the MSSM, including correlations between $b\to s\nu\bar\nu$ and $s\to d\nu\bar\nu$ transitions, in section~\ref{NPMSSM}. For an analysis in a single universal extra dimension see ref.~\cite{Colangelo:2006vm}.

Since the neutrinos originating from $b\to s\nu\bar\nu$ decays cannot be detected experimentally but only manifest themselves as missing energy, the actual processes being measured are $B\to(K,K^*,X_s)+\slashed E$.
Therefore, New Physics can enter the observables not only through a modification of the Wilson coefficients,
but also 
through invisible decays to unknown particles overlapping with the $b\to s\nu\bar\nu$ decays. We discuss one such model, in which the neutrinos are replaced by neutral scalars, in section~\ref{sec:scalars}. Similar studies in the context of the NMSSM and unparticle physics were presented in \cite{Hiller:2004ii,Aliev:2007gr}.

\begin{table}[tb]
\centering
\renewcommand{\arraystretch}{1.1}
\begin{tabular}{|lll|lll|}
\hline
Parameter & Value & Ref. & Parameter & Value & Ref.\\
\hline\hline
$m_b^{1S}$ & $(4.68 \pm 0.03)$ GeV & \cite{Hoang:2000fm,Bauer:2004ve} & $\lambda$  & 0.2255(7) & \cite{Antonelli:2008jg} \\
$m_s(2~\text{GeV})$ & 0.1 GeV & \cite{Amsler:2008zzb} & $|V_{cb}|$  & $(4.13\pm0.05) \times 10^{-2}$ & \cite{Bona:2006ah} \\
$m_t(m_t)$ & $(162.3\pm1.2)$ GeV & \cite{:2008vn} & $\bar\rho$  & $0.154\pm0.022$ & \cite{Bona:2006ah} \\
$\tau_{B^+}$ & 1.638 ps & \cite{Amsler:2008zzb} & $\bar\eta$  & $0.342\pm0.014$ & \cite{Bona:2006ah} \\
$\tau_{B^0}$ & 1.530 ps & \cite{Amsler:2008zzb} & $\lambda_1$ &  $(-0.27 \pm 0.04)$ GeV$^2$ & \cite{Bauer:2004ve}\\
 &  &  & $\lambda_2$ &  $(0.12 \pm 0.01)$ GeV$^2$ & \cite{Amsler:2008zzb}\\
\hline
\end{tabular}
\renewcommand{\arraystretch}{1}
\caption[]{\small Parameters used in the numerical analysis. $\lambda_{1,2}$ are the HQET parameters needed for the evaluation of the $\Lambda^2/m_b^2$ corrections to $\text{BR}(B\to X_s\nu\bar\nu)$ \cite{Grossman:1995gt}.}\label{tab:par}
\end{table}

\subsection{Standard Model}\label{sec:SM}

Neither the inclusive nor the two exclusive $b\to s\nu\bar\nu$ decay modes have been observed in experiment so far. However, experimental upper bounds on the branching ratios have been set by the BaBar, Belle and ALEPH collaborations. We summarize them in table~\ref{tab:exp}, together with our predictions for their SM values.
\begin{table}[tb]
\centering
\renewcommand{\arraystretch}{1.5}
\begin{tabular}{|l|l|l|}
\hline
Observable  &  Our SM prediction &  Experiment \\
\hline\hline
$\text{BR}(B \to K^* \nu\bar\nu)$ & $( 6.8^{+1.0}_{-1.1} ) \times 10^{-6}$ & $< 80 \times 10^{-6}$ \cite{:2008fr} \\
$\text{BR}(B^+ \to K^+ \nu\bar\nu)$   & $( 4.5 \pm 0.7 ) \times 10^{-6}$ & $< 14 \times 10^{-6}$ \cite{:2007zk} \\
$\text{BR}(B \to X_s \nu\bar\nu)$ & $( 2.7\pm0.2 ) \times 10^{-5}$ & $< 64 \times 10^{-5}$ \cite{Barate:2000rc} \\
$\langle F_L(B \to K^* \nu\bar\nu) \rangle$ & $0.54 \pm 0.01$ & --  \\
\hline
\end{tabular}
\renewcommand{\arraystretch}{1}
\caption[]{\small SM predictions and experimental bounds (all at the 90\% C.L.) for the four $b\to s\nu\bar\nu$ observables.}\label{tab:exp}
\end{table}

\begin{figure}[tb]
\centering
\includegraphics[width=\textwidth]{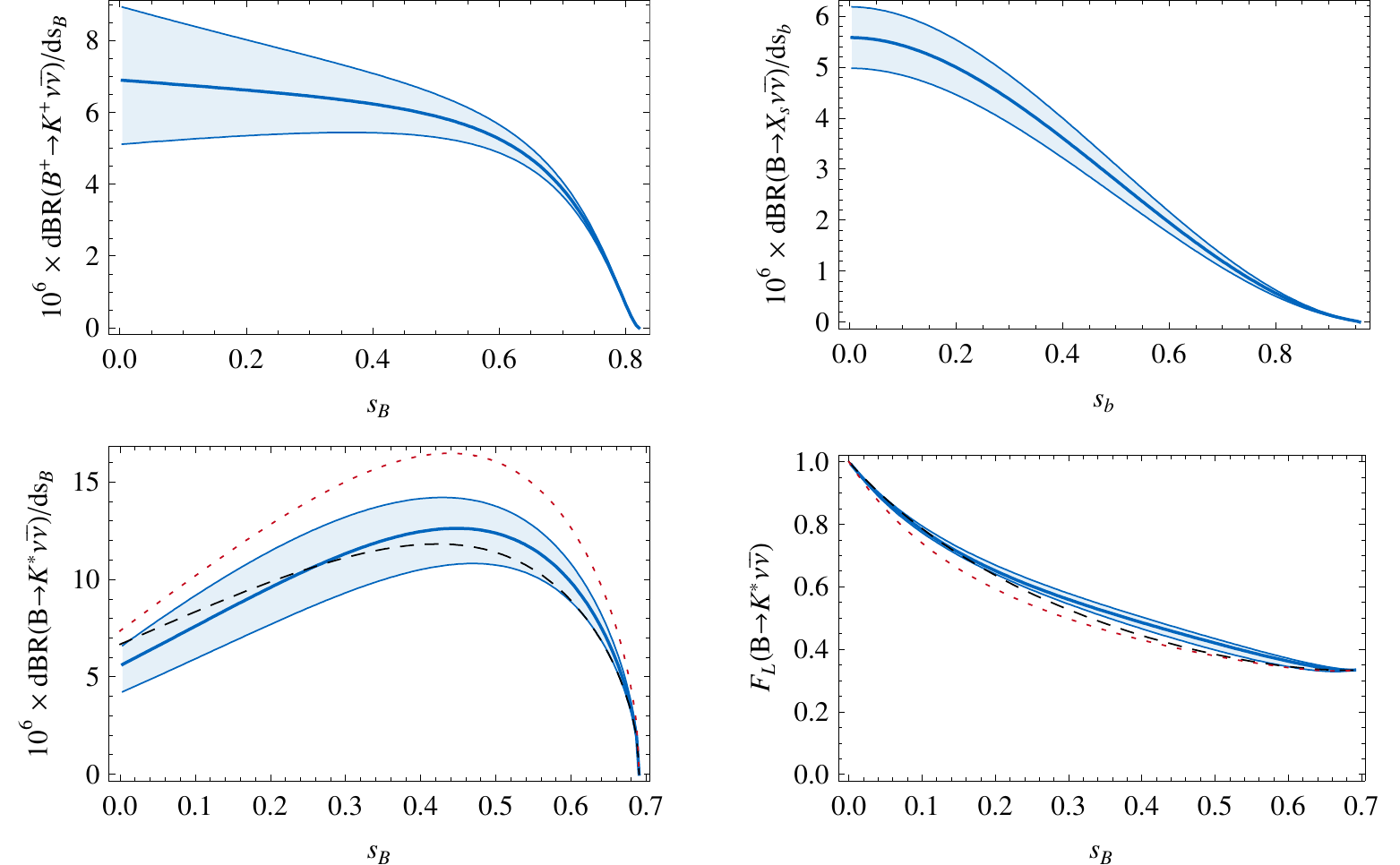}
\caption[]{\small Dependence of the four $b\to s\nu\bar\nu$ observables on the normalized neutrino invariant masses squared $s_{b,B}$ within the SM. The error bands reflect the theoretical uncertainties. In the lower plots, the black dashed lines and dotted red lines are the results based on the form factor sets $B$ and $C$, respectively. See the text for more details.}
\label{fig:SM}
\end{figure}
In figure~\ref{fig:SM}, we show our SM predictions for the differential branching ratios of all three decays and for $F_L(s_B)$. Concerning the observable $F_L(s_B)$, it is interesting to note that the value $F_L(0)=1$ is due to helicity conservation, forcing the $B$ meson to decay into a longitudinal $K^*$. The kinematical endpoint at $s_B=(1-\widetilde{m}_{K^*})^2$ corresponds to the case of zero spatial momentum of the $K^*$ in the $B$ restframe. The absence of a preferential direction at this point explains the value of $1/3$ as the ratio of the single longitudinal polarization state to the total number of 3 states.

Next we want to illustrate briefly the dependence of the SM prediction on the choice of $B \to K^*$ form factors. To this end we plot in figure~\ref{fig:SM} in addition to the results of our main set of form factors (set $A$) the observables for two older sets: set $B$ from ref. \cite{Melikhov:1996du} and set $C$ from ref. \cite{Ball:2004rg}. Since a discussion of the technical differences and the error estimates of the other sets is beyond the scope of this work, we give here only the central values. While the prediction for the differential branching ratio is similar for sets $A$ and $B$, there is a difference of about 25\% relative to the results obtained from set $C$. This reflects a quite general offset of the relevant form factors ($V$, $A_1$, $A_2$) of set $C$ relative to the other sets, which is due to our differing strategies in the normalization of the form factors. As discussed in \cite{Altmannshofer:2008dz}, our form factors are normalized 
such that the tensor form factor $T_1(q^2=0)$ reproduces the experimental value of $\text{BR}(B \to K^* \gamma)$, which implies $T_1(0)=0.267\pm0.018$ \cite{Ball:2006eu}. The resulting uncertainty in the overall normalization of the form factors of about 7\% is taken into account in our uncertainty estimates.
We emphasize however that this effect of the differing normalizations is absent in $F_L$, since overall factors cancel in this ratio. For completeness, we give the branching ratios obtained by using the two older sets of form factors together with our values for the parameters as in table~\ref{tab:par}: $\text{BR}(B \to K^* \nu\bar\nu)_B = 6.7 \times 10^{-6}$, $\text{BR}(B \to K^* \nu\bar\nu)_C = 8.9 \times 10^{-6}$. We note that both these values and our prediction for $\text{BR}(B \to K^* \nu\bar\nu)$ are lower than the ones present in the literature \cite{Buchalla:2000sk,Hurth:2008jc}.

The estimates of the theoretical uncertainties in table~\ref{tab:exp} and the error bands in figure~\ref{fig:SM} include the uncertainties due to the form factors in the case of the exclusive decays and the uncertainties of the CKM elements as listed in table~\ref{tab:par} as well as the uncertainty in the SM Wilson coefficient as given in eq.~(\ref{eq:CLSM}), for all decays.

For the inclusive decay, the uncertainty is dominated by the theory error of $m_b^{1S}$. For the branching ratio prediction, we took into account the $O(\Lambda^2/m_b^2)$ corrections and the corresponding errors of $\lambda_{1,2}$ as indicated in table~\ref{tab:par}. To be conservative, we assume an additional uncertainty of the inclusive branching ratio of 5\% to account for neglected higher order corrections. For the inclusive dineutrino mass spectrum in figure~\ref{fig:SM}, we omitted the $O(\Lambda^2/m_b^2)$ corrections, since they become singular at the kinematical endpoint. Therefore, in order to be on the conservative side and bearring in mind that local quantities are harder to estimate we increased the additional error on the dineutrino mass spectrum to 10\%. Such problems do not arise in the prediction of a global quantity as the branching ratio.

Finally, we added all the individual uncertainties in quadrature.

\subsection{Model-independent constraints on Wilson coefficients}\label{sec:indep}

\begin{figure}[tb]
\centering
\includegraphics[width=0.45\textwidth]{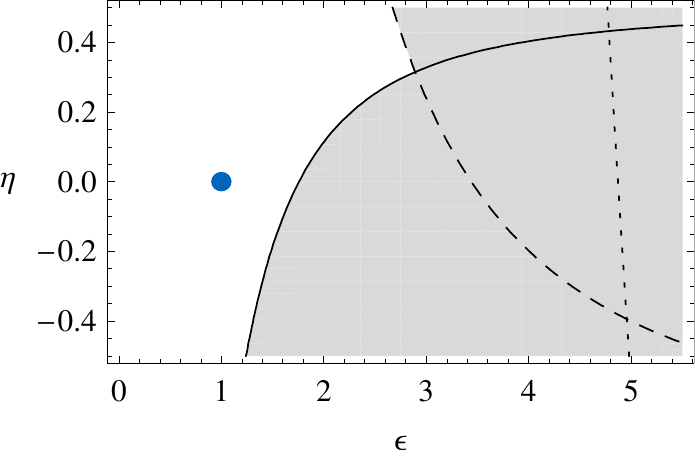}
\caption[]{\small Existing experimental constraints on $\epsilon$ and $\eta$. Dashed line: constraint from $\text{BR}(B \to K^* \nu\bar\nu)$, solid line: constraint from $\text{BR}(B \to K \nu\bar\nu)$, dotted line: constraint from $\text{BR}(B \to X_s \nu\bar\nu)$. The shaded area is ruled out experimentally at the 90\% confidence level. The blue circle represents the SM point. }
\label{fig:constraints}
\end{figure}

The four observables accessible in the three different $b\to s\nu\bar\nu$ decays are dependent on the two in principle complex Wilson coefficients $C^\nu_L$ and $C^\nu_R$. However, only two combinations of these complex quantities enter the formulae given in section~\ref{sec:obs} and are thus observable. These are \cite{Grossman:1995gt,Melikhov:1998ug} 
\begin{equation}  \label{eq:epsetadef}
 \epsilon = \frac{\sqrt{ |C^\nu_L|^2 + |C^\nu_R|^2}}{|(C^\nu_L)^\text{SM}|}~, \qquad
 \eta = \frac{-\text{Re}\left(C^\nu_L C_R^{\nu *}\right)}{|C^\nu_L|^2 + |C^\nu_R|^2}~,
\end{equation}
such that $\eta$ lies in the range $[-\frac{1}{2},\frac{1}{2}]$.
The observables discussed in section~\ref{sec:obs} can be expressed in terms of $\epsilon$ and $\eta$ as follows
\begin{align}
\label{eq:epseta-BKsnn}
 \text{BR}(B \to K^* \nu\bar\nu) & = 6.8 \times 10^{-6} \, (1 + 1.31 \,\eta)\epsilon^2~, \\
\label{eq:epseta-BKnn}
 \text{BR}(B \to K \nu\bar\nu)   & = 4.5 \times 10^{-6} \, (1 - 2\,\eta)\epsilon^2~, \\
\label{eq:epseta-BXsnn}
 \text{BR}(B \to X_s \nu\bar\nu) & = 2.7 \times 10^{-5} \, (1 + 0.09 \,\eta)\epsilon^2~, \\
\label{eq:epseta-FL}
 \langle F_L \rangle             & = 0.54 \, \frac{(1 + 2 \,\eta)}{(1 + 1.31 \,\eta)}~.
\end{align}
As $\epsilon$ and $\eta$ can be calculated in any model by means of eq. (\ref{eq:epsetadef}), these four expressions can be considered as fundamental formulae for any phenomenological analysis of the decays in question.
The experimental bounds on the branching ratios, cf. table~\ref{tab:exp}, can then be translated to excluded areas in the $\epsilon$-$\eta$-plane, see figure~\ref{fig:constraints}, where the SM corresponds to $(\epsilon,\eta)=(1,0)$. We observe that the exclusive decays are presently more constraining than the inclusive one.

\begin{figure}[tb]
\centering
\includegraphics[width=0.45\textwidth]{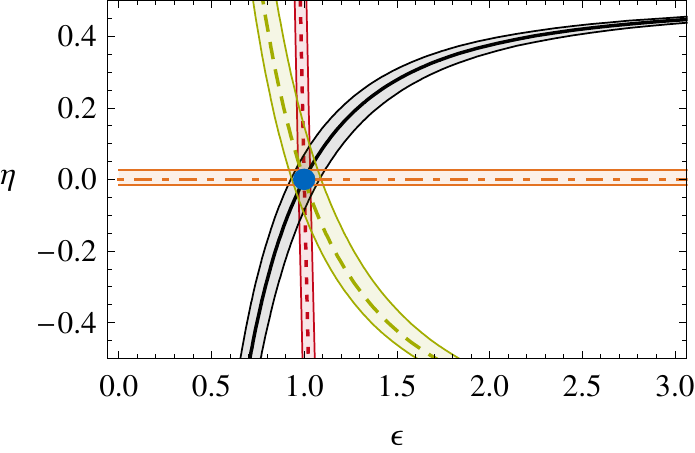}\qquad
\includegraphics[width=0.45\textwidth]{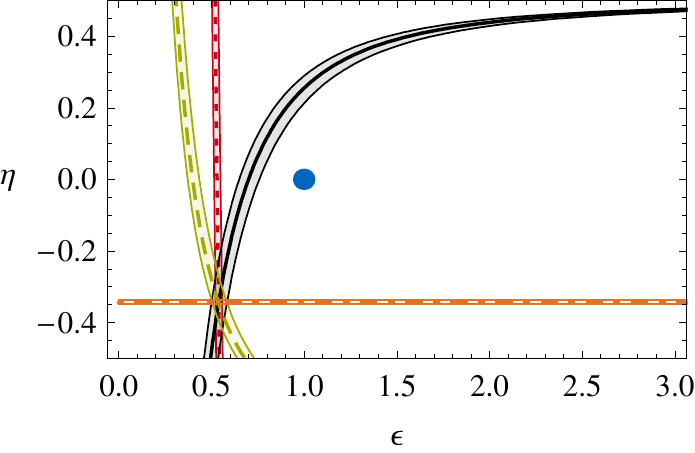}
\caption[]{\small Hypothetical constraints on the $\epsilon$-$\eta$-plane, assuming all four observables have been measured with infinite precision. The error bands reflect the theoretical uncertainty as described in section~\ref{sec:SM}.
The green band (dashed line) represents $\text{BR}(B \to K^* \nu\bar\nu)$, the black band (solid line) $\text{BR}(B \to K \nu\bar\nu)$, the red band (dotted line) $\text{BR}(B \to X_s \nu\bar\nu)$ and the orange band (dot-dashed line) $\langle F_L \rangle$.
Left: SM values for the Wilson coefficients, right: assuming $C^\nu_L=0.5(C^\nu_L)^\text{SM}$ and $C^\nu_R=0.2(C^\nu_L)^\text{SM}$. The blue circle represents the SM point. }
\label{fig:exp-hypo}
\end{figure}

Since the four observables depend on only two parameters, a measurement of all of them would overconstrain the resulting $(\epsilon,\eta)$ point. To illustrate the theoretical cleanliness of the various observables, we show in figure~\ref{fig:exp-hypo} the combined constraints after hypothetical measurements with infinite precision, first assuming the SM and then for a toy NP example.

A special role is played by the observable $\langle F_L \rangle$: since it only depends on $\eta$, cf. eq. (\ref{eq:epseta-FL}), it leads to a horizontal line in the $\epsilon$-$\eta$ plane. Although a similar constraint could be obtained by dividing two of the branching ratios to cancel the common factor of $\epsilon^2$, the use of $\langle F_L \rangle$ is theoretically much cleaner since in this case, the hadronic uncertainties cancel, while they would add up when using the branching ratios.

\begin{figure}[t]
\centering
\includegraphics[width=0.99\textwidth]{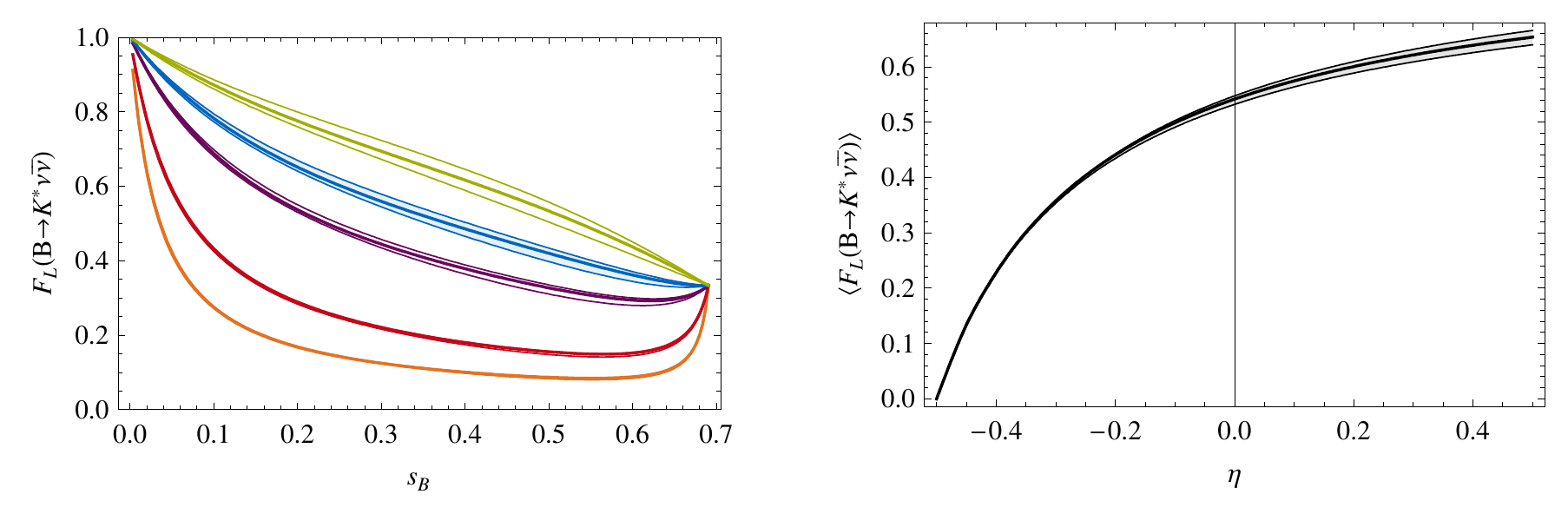}
\caption[]{\small Left: $F_L(s_B)$ for different values of $\eta$, from top to bottom: $\eta=0.5, 0, -0.2, -0.4, -0.45$.
Right: Dependence of the $s_B$-integrated $\langle F_L \rangle$ on $\eta$.}
\label{fig:FL-NP}
\end{figure}
In the right-hand panel of figure \ref{fig:FL-NP}, we show the value of $\langle F_L \rangle$ as a function of $\eta$. Especially for negative $\eta$, $\langle F_L \rangle$ constitutes a very clean observable to probe the value of $\eta$.

Another interesting point about $F_L$ is that, since it only depends on $\eta$, the distribution $F_L(s_B)$ is universal for all models in which one of the Wilson coefficients $C^\nu_{L,R}$ vanishes, such as in the SM and models with constrained minimal flavour violation (CMFV) \cite{Hurth:2008jc,Buras:2000dm,D'Ambrosio:2002ex}. In the left-hand panel of figure \ref{fig:FL-NP}, we plot $F_L(s_B)$ in the kinematically allowed range of $s_B$ for several values of $\eta$. The blue curve is the universal curve for $\eta=0$. Every experimentally observed deviation from this curve signals clearly the presence of right-handed currents as left-handed currents are non-vanishing.

\subsection[Modified $Z^{(\prime)}$ penguins]{Modified \boldmath$Z^{(\prime)}$ penguins}\label{sec:penguins}

In many models beyond the SM, NP effects in the Wilson coefficients $C^\nu_{L,R}$ are dominated by $Z$ penguins. This can be discussed model-independently by assuming an effective flavour violating $\bar bsZ$ coupling \cite{Buchalla:2000sk}, which will not only modify the Wilson coefficients $C^\nu_{L,R}$, but also the Wilson coefficients $C_{9,10}^{(\prime)}$ of the semi-leptonic operators governing $b\to s\ell^+\ell^-$ transitions. Therefore, interesting correlations between these processes and the $b\to s\nu\bar\nu$ transitions are to be expected in this scenario.

\subsubsection{Effective Lagrangian}
The flavour violating $\bar bsZ$ coupling can be parametrized in terms of the effective Lagrangian \cite{Buchalla:2000sk}
\begin{equation}
 \mathcal L^{\bar bsZ}_\text{eff} = \frac{G_F}{\sqrt{2}} \frac{e}{\pi^2}m_Z^2 c_w s_w V^*_{tb}V_{ts} \; Z^\mu \left(
Z_L \; \bar b \gamma_\mu P_L  s
+ Z_R \; \bar b \gamma_\mu P_R  s
\right)~,
\label{eq:LbsZ}
\end{equation}
with $s_w = \sin\theta_w$ and $c_w = \cos\theta_w$. In the SM, the right-handed coupling is negligible, while $Z_L =  C_0(x_t)/s_w^2$. The function $C_0$ can be found e.g. in \cite{Buchalla:1995vs}.
In models with CMFV, $Z_L$ is a real function of the model parameters and $Z_R$ is strongly suppressed, while in general NP models $Z_L$ and $Z_R$ can be arbitrary complex couplings.

It should be remarked that the $Z$ penguins are generally gauge dependent. In the SM, this gauge dependence is rather weak as it enters only in non-leading terms in $m_t$ and is cancelled through box diagrams and photon penguin diagrams. As the latter diagrams receive subdominant contributions in most extensions of the SM with respect to NP contributions to $Z$ penguins, we expect that the gauge dependence of NP contributions to $Z_{L,R}$ is also very weak and it is a very good approximation to parametrize the NP contributions by the modifications of $Z_{L,R}$ only \cite{Buras:1998ed}. Arguments for NP modifying dominantly $Z$ penguin contributions are given in \cite{Buras:1999da}.

\subsubsection[Constraints on modified $Z$ penguins]{Constraints on modified \boldmath$Z$ penguins}
The impact of NP effects in the $\bar bsZ$ couplings $Z_{L,R}$ on the Wilson coefficients is\footnote{%
Our convention for the Wilson coefficients $C^{(\prime)}_{9,10}$ is such that they equal the quantities $C_{9,10}^{(\prime)\text{eff}}$ of ref. \cite{Altmannshofer:2008dz}.
} \cite{Buchalla:2000sk}
\begin{align}
C^\nu_L &= (C^\nu_L)^\text{SM} - Z_L^\text{NP}~, & C^\nu_R &=  - Z_R~,
\label{eq:WCs-bsZ-1}\\
C_{10} &= C_{10}^\text{SM}  - Z_L^\text{NP}~, & C'_{10} &=  - Z_R~,\\
C_9 &= C_9^\text{SM}   +  Z_L^\text{NP}(1-4s_w^2)~, & C'_9 &=  Z_R (1-4s_w^2)~.
\label{eq:WCs-bsZ-3}
\end{align}
The contributions to $C_9^{(\prime)}$ are strongly suppressed by the small vector coupling of the $Z$ to charged leptons $(1-4s_w^2)\approx0.08$.

The most stringent constraint on $Z_{L,R}^\text{NP}$ comes from the measurement of the branching ratio of the inclusive decay $B\to X_s\ell^+\ell^-$, which reads in the low-$q^2$ region, $1\,\text{GeV}^2<q^2<6\,\text{GeV}^2$ \cite{Aubert:2004it,Iwasaki:2005sy},
\begin{equation}  
\text{BR}(B\to X_s\ell^+\ell^-)_\text{exp.} = (1.60 \pm 0.51) \times 10^{-6}~.
\label{eq:BRBXsellell}
\end{equation}
Assuming that NP contributions enter exclusively through modified $Z$ penguins, which we will assume throughout this section, this can be translated into a bound on the flavour-changing $Z$ couplings,
\begin{equation}
4.3 < |Z_L|^2+|Z_R|^2 < 28.8
\label{eq:ZLR-constraint}
\end{equation}
at the $1\sigma$ level. An additional (currently weaker) constraint arises from the experimental upper bound on the branching ratio of $B_s\to\mu^+\mu^-$ \cite{:2007kv},
\begin{equation}  
\text{BR}(B_s\to\mu^+\mu^-)_\text{exp.} < 5.8 \times 10^{-8} ~\text{at 95\% C.L.}~,
\end{equation}
leading to
\begin{equation}
|Z_L - Z_R|^2 < 261~,
\end{equation}
again assuming that scalar or pseudoscalar operator contributions to $B_s\to\mu^+\mu^-$ are negligible.

The couplings $Z_{L,R}$ will also contribute to $B_s$-$\bar B_s$ mixing via double $Z$ penguin diagrams, which contribute to the amplitude a term
\begin{equation}
\frac{%
 \langle B_s | \mathcal H | \bar B_s \rangle^{\bar bsZ} }{%
 \langle B_s | \mathcal H | \bar B_s \rangle^\text{SM} }
	= \frac{4\alpha s_w^2}{\pi S_0(x_t)} ( Z_L^2 + x Z_L Z_R + Z_R^2 )~,
\label{eq:DMs-ZLR}
\end{equation}
where the function $S_0$ can be found e.g. in \cite{Buchalla:1995vs} and $x$ is a hadronic parameter containing the ratio of hadronic matrix elements of the respective $\Delta B = 2$ operators. With LO QCD running for the involved operators, we find 
\begin{equation}
x = - \frac{m_{B_s}^2}{(m_b+m_s)^2} \frac{B^{(s)}_5}{B^{(s)}_1} \left( \frac{\alpha_s(m_Z)}{\alpha_s(m_b)}\right)^{-\frac{3}{23}} \simeq - 3.5~,
\end{equation}
where for the numerical evaluation we used the B-parameters $B^{(s)}_1$ and $B^{(s)}_5$ in the $\overline{\text{MS}}$ scheme from \cite{Becirevic:2001xt}.
The amplitude is usually parametrized as
\begin{equation}
\langle B_s | \mathcal H | \bar B_s \rangle = \frac{\Delta M_s}{2}\; e^{2i (\phi_{B_s} + \beta_s)} ~.
\end{equation}
The mass difference has been measured to be \cite{Abulencia:2006ze}
\begin{equation}
(\Delta M_s)_\text{exp.} = (17.77\pm0.12)\,\text{ps}^{-1} ~,
\end{equation}
however, the theory prediction is afflicted with an uncertainty of roughly 30\% due to uncertainties in hadronic parameters. While the $B_s$ mixing phase predicted by the SM is tiny, $\beta_s\approx1^\circ$, recent Tevatron data seem to indicate the presence of a sizable phase $\phi_{B_s}$ \cite{Bona:2008jn,Lenz:2008dp,Brooijmans:2008nt,Lenz:2006hd,Ball:2006xx}.

\begin{figure}[tb]
\centering
\includegraphics[width=0.57\textwidth]{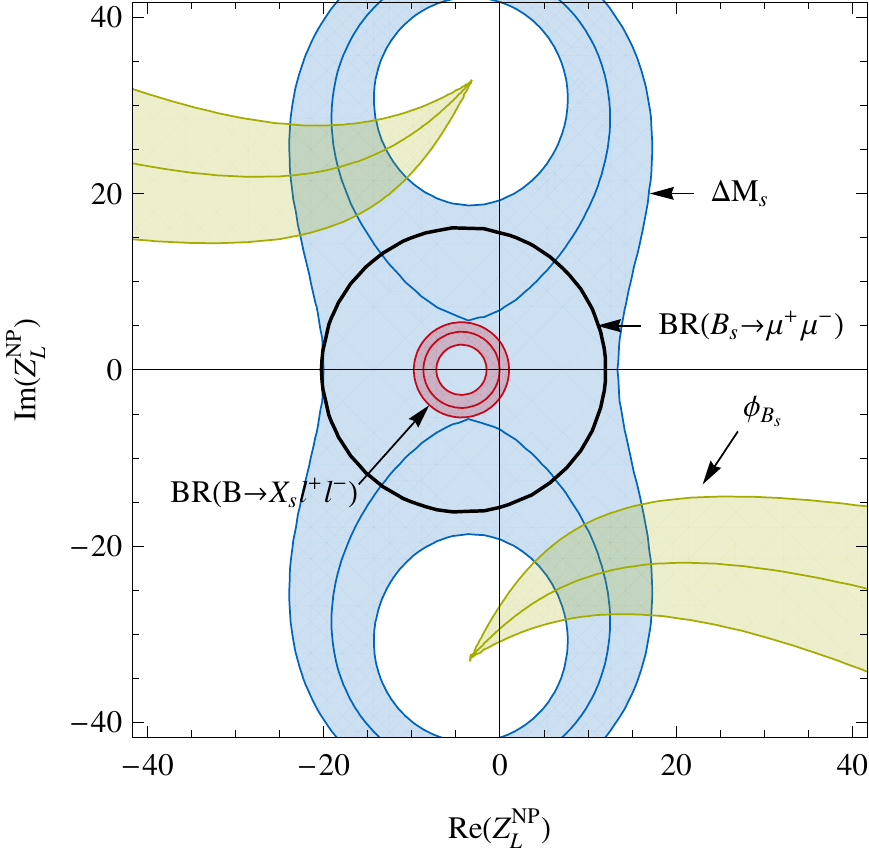}
\caption[]{\small Constraints on the real and imaginary parts of $Z_L^\text{NP}$ coming from $\Delta M_s$ (blue, assuming 30\% theory uncertainty), $\text{BR}(B\to X_s\ell^+\ell^-)$ (red) and $\text{BR}(B_s\to\mu^+\mu^-)$ (black) assuming $Z_R=0$. The green lines correspond to values of the $B_s$ mixing phase $\phi_{B_s}=-11^\circ$, $-19^\circ$ and $-27^\circ$, respectively \cite{Bona:2008jn}.}
\label{fig:ZL}
\end{figure}

In principle, large complex $\bar bsZ$ couplings $Z_{L,R}$ could give rise to a such a phase. However, taking into account the constraint in eq. (\ref{eq:ZLR-constraint}), the double penguin contribution is too small to generate a sizable phase. We visualize the constraints from $B\to X_s\ell^+\ell^-$, $B_s\to\mu^+\mu^-$ and from $B_s$ mixing in figure~\ref{fig:ZL} for the case $Z_R=0$. In the general case of nonzero and complex $Z_L$ and $Z_R$, the correlation is more complicated 
(e.g., for $Z_L=Z_R$ the constraint from $B_s\to\mu^+\mu^-$ disappears) but 
we find that it is never possible to bring the stringent constraint from 
$B\to X_s\ell^+\ell^-$ into agreement with a large $B_s$ mixing phase\footnote{%
As pointed out in \cite{Bobeth:2008ij}, the experimental indication of a SM-like sign of the forward-backward asymmetry of $B\to K^*\ell^+\ell^-$ in the high-$q^2$ region \cite{Ishikawa:2006fh,:2008ju} puts additional constraints on $C_{10}^\text{NP}$ (and thus on $Z_L^\text{NP}$), further strengthening this conclusion.
}.

\begin{figure}[tb]
\centering
\includegraphics[width=\textwidth]{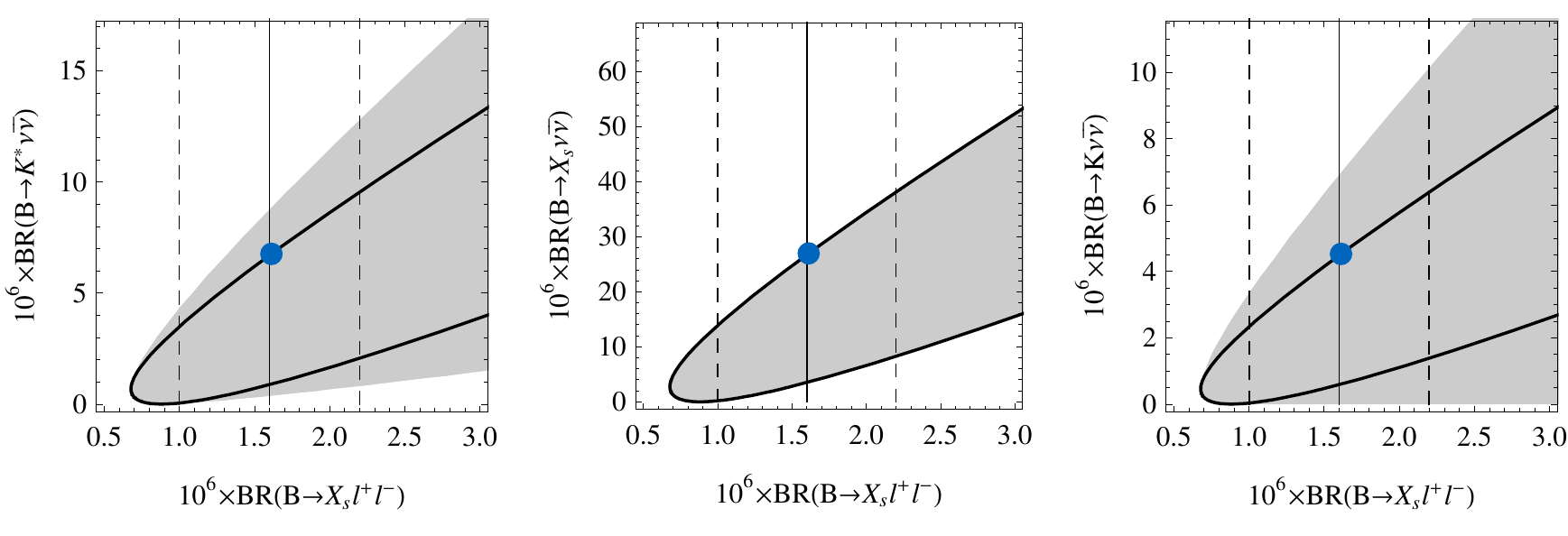}
\caption[]{\small Correlations between $b\to s\nu\bar\nu$ branching ratios and $\text{BR}(B\to X_s\ell^+\ell^-)$. The black curves correspond to $Z_R=0$ and real $Z_L$; The shaded areas are accessible for arbitrary $Z_{L,R}$; The blue dots represent the SM. The solid and dashed vertical lines correspond to the experimental central value and $1\sigma$ error, respectively, of $\text{BR}(B\to X_s\ell^+\ell^-)$.}
\label{fig:nunu-vs-ll}
\end{figure}

In figure \ref{fig:nunu-vs-ll}, we show the correlation between the three $b\to s\nu\bar\nu$ branching ratios and $\text{BR}(B\to X_s\ell^+\ell^-)$. Assuming $Z_R=0$ and $Z_L$ real, which holds in CMFV models, there are clear correlations, indicated as black curves, between the neutrino modes and the charged lepton mode. In the general case of arbitrary and complex $Z_{L,R}$, the entire shaded areas are accessible. It is interesting to note, however, that in all three $b\to s\nu\bar\nu$ decay modes, an enhancement of the branching ratio by more than a factor of two with respect to the SM is excluded by the measurement of $\text{BR}(B\to X_s\ell^+\ell^-)$ in eq. (\ref{eq:BRBXsellell}). By construction, this statement is valid for all models in which NP contributions to $b\to s\nu\bar\nu$ and $b\to s\ell^+\ell^-$ processes enter dominantly through flavour-changing $Z$ penguins.

\subsubsection[Flavour violating $Z'$ couplings]{Flavour violating \boldmath $Z'$ couplings}
One way to circumvent this constraint is by replacing the $Z$ boson in the above considerations by the $Z'$ gauge boson of an additional $U(1)'$ symmetry, i.e. assuming an SM-like $\bar bsZ$ coupling but a flavour violating $\bar bsZ'$ coupling. Then, instead of eq. (\ref{eq:LbsZ}), one has
\begin{equation}
 \mathcal L^{\bar bsZ'}_\text{eff} = \frac{G_F}{\sqrt{2}} \frac{e}{\pi^2}m_{Z'}^2 c_w s_w V^*_{tb}V_{ts} \; Z'^\mu \left(
Z'_L \; \bar b \gamma_\mu P_L  s
+ Z'_R \; \bar b \gamma_\mu P_R  s
\right)~.
\end{equation}
Such couplings can arise either as effective couplings induced by loop effects of particles charged under the $U(1)'$, or even at tree level in the case of generation non-universal $U(1)'$ charges of the quarks \cite{Langacker:2000ju}.
In this setup, the analogues to eqs. (\ref{eq:WCs-bsZ-1})--(\ref{eq:WCs-bsZ-3}) read
\begin{align}
C^\nu_L &= (C^\nu_L)^\text{SM} - \frac{g^{\prime\nu}_V}{2} Z'_L~, & C^\nu_R &= - \frac{g^{\prime\nu}_V}{2} Z'_R~, \\
C_{10} &= C_{10}^\text{SM} + \frac{g^{\prime\ell}_A}{2} Z'_L~, & C'_{10} &= + \frac{g^{\prime\ell}_A}{2} Z'_R~,\\
C_9 &= C_9^\text{SM} - \frac{g^{\prime\ell}_V}{2} Z'_L~, & C'_9 &=- \frac{g^{\prime\ell}_V}{2} Z'_R~,
\end{align}
where the couplings $g^{\prime\nu,\ell}_{V,A}$ denote the vector and axial vector couplings of the $Z'$ to neutrinos and charged leptons, respectively. These couplings are given by the $U(1)'$ charges of the respective fields and are arbitrary -- apart from anomaly constraints, which can however always be fulfilled by adjusting the quark $U(1)'$ charges and/or adding new, exotic fermions.

The contribution to the $B_s$ mixing amplitude, on the other hand, is independent of the $g'$ couplings and is simply given by eq.~(\ref{eq:DMs-ZLR}) after the replacements $Z_{L,R}\to Z'_{L,R}$. Therefore, in a general $Z'$ model, by choosing small or zero $U(1)'$ charges for the charged leptons it is possible in principle to completely suppress the NP contributions to $b\to s\ell^+\ell^-$ as well as $B_s\to\ell^+\ell^-$ decays, while it is at the same time possible to obtain a strong enhancement of $b\to s\nu\bar\nu$ modes and/or a sizable, potentially complex, contribution to the $B_s$ mixing amplitude.

\subsection{Littlest Higgs with T-Parity (LHT)}  \label{NPLHT}
Right-handed currents are absent or suppressed in most NP models. One example is the Littlest Higgs model with T-parity, where $C^\nu_R$ is negligible by construction and NP effects in $C^\nu_L$ are rather small \cite{Blanke:2006eb}. A scan over the parameter space shows that $(C_L^\nu)^\text{NP}$ typically amounts to 10\% of the SM value if experimental constraints from other flavour physics observables are imposed. Consequently, it will be difficult to distinguish this model from the SM on the basis of the decays considered here.

\boldmath
\subsection{RS model with custodial protection of left-handed $Z$ couplings} \label{sec:WED}
\unboldmath
Recently the decays $B\to K^*\nu\bar\nu$, $B\to K\nu\bar\nu$ and 
$B\to X_{s,d}\nu\bar\nu$ have been analyzed in a Randall-Sundrum model
with a custodial protection of the {\it left-handed}
 $Z$ couplings to down-quarks
\cite{Blanke:2008yr}. In this model the NP contributions to the decays in 
questions are
dominated then by tree level $Z$ boson exchanges governed by {\it
right-handed} couplings to down-quarks. In spite of $C_R^\nu$ being
non-vanishing in this model, the deviations from the SM for the 
three decays considered here are found to be even smaller than in the LHT
model. Interestingly, when the custodial protection of left-handed $Z$
couplings is removed, NP effects in $b\to s\nu\bar\nu$ transitions can 
be enhanced relative to the SM by as much as a factor of three which is not
possible in the LHT model and in several NP scenarios considered here. However,
in such a scenario also a strong violation of the experimental constraint
on the $Zb_L\bar b_L$ coupling is predicted and a consistent analysis should take
into account also electroweak precision observables.

\subsection{Minimal Supersymmetric Standard Model} \label{NPMSSM}
\subsubsection{General considerations}

In the Minimal Supersymmetric Standard Model (MSSM) with a generic flavour violating soft sector there are various new contributions to the $b \to s \nu \bar \nu$ transition \cite{Bertolini:1990if,Goto:1996dh,Bobeth:2001jm,Buras:2004qb,Yamada07} and one might expect that large effects are possible. However, once the existing constraints coming from other flavour changing processes are applied, the effects in $C_L^\nu$ and particularly in $C_R^\nu$ turn out to be quite limited in the MSSM \cite{Yamada07,Grossman:1995gt}.

While neutralino contributions are generally expected to be small, gluino contributions to both $C_L^\nu$ and $C_R^\nu$ are highly constrained by the $b \to s \gamma$ decay and have only negligible impact. Charged Higgs contributions to $C_L^\nu$ scale as $1/\tan^2\beta$ and even for low values of $\tan\beta$ they play only a marginal role. Concerning the charged Higgs contributions to the right handed coefficient $C_R^\nu$, at the leading order, they are proportional to $m_s m_b \tan^2\beta$ and therefore negligible even for large values of $\tan\beta$. On the other hand, non-holomorphic corrections to the Higgs couplings can enhance this contribution and can lead to important effects in the large $\tan\beta$ regime, as is well known in the case of $s \to d \nu\bar\nu$ transitions \cite{Isidori:2006jh}. In the case of $b \to s \nu\bar\nu$ transitions however, we confirm the expectation of \cite{Yamada07} that the upper bound on the branching ratio of the rare decay $B_s \to \mu^+ \mu^-$ sets strong limits on this contribution, that then also turns out to be negligible.

Turning to chargino contributions to the right handed coefficient $C_R^\nu$, at the leading order they are also suppressed by $m_s m_b \tan^2\beta$, as the Higgs contributions are, and therefore negligible. One is then left with the chargino contributions to the left handed coefficient $C_L^\nu$ that are the only ones where sizable effects are still possible. Largest effects can be generated by a $Z$ penguin with a $(\delta_u^{RL})_{32}$ mass insertion \cite{Lunghi:1999uk,Buchalla:2000sk,Ali:2002jg}, that is not strongly constrained by existing data \cite{Misiak:1997ei,Colangelo:1998pm,Lunghi:1999uk,Xiao:2006gu}.

\begin{figure}[tb]
\centering
\includegraphics[width=0.66\textwidth]{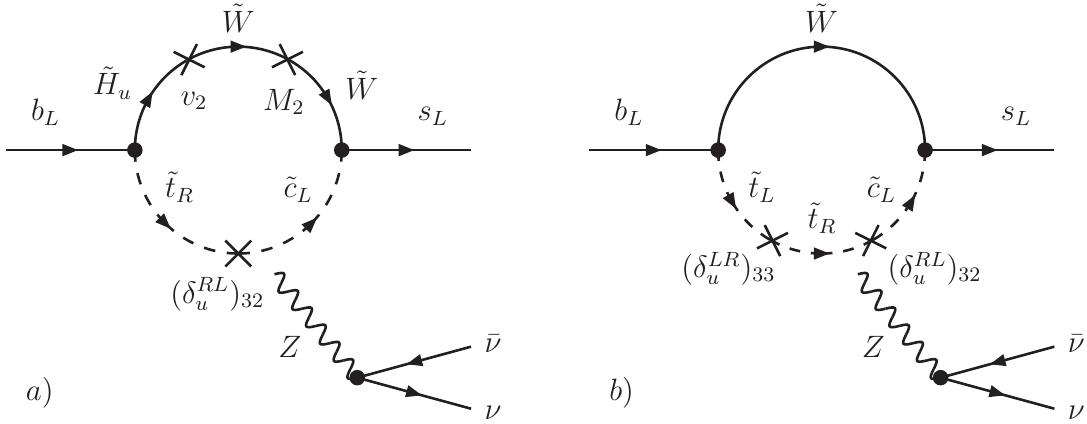}
\caption[]{\small Dominant chargino contributions to the Wilson coefficient $C_L^\nu$ in the mass insertion approximation.}
\label{fig:Bdiagrams}
\end{figure}

The $Z$ penguin diagrams giving that contribution are shown in figure~\ref{fig:Bdiagrams} and the corresponding analytical expression in the mass insertion approximation reads\footnote{In our numerical analysis, we work with mass eigenstates and include the complete set of SUSY contributions as given in \cite{Buras:2004qb}.}

\begin{equation}\label{eq:CLnu}
(C_L^\nu)^{\tilde\chi^\pm} \simeq - \frac{1}{s_w^2} \frac{V_{cs}^*}{V_{ts}^*} (\delta_u^{RL})_{32} \left[ \frac{m_t A_t}{8 \tilde m^2} f_1(x_2) - \frac{m_t M_2}{4 \tilde m^2} f_2(x_\mu,x_2) \right]~,
\end{equation}
where $M_2$ is the Wino mass, $A_t$ is the trilinear coupling of the stop and for simplicity we assumed that the masses of the left and right handed up-type squarks have a common value $m_{\tilde Q}^2 = m_{\tilde U}^2 = \tilde m^2$.
Our conventions for the up squark mass is such that $(M^2_{\tilde U})^{LR}_{33} = - m_t (A_t + \mu^* \cot\beta)$ and $(M^2_{\tilde U})^{RL}_{32} = (\delta_u^{RL})_{32} m_{\tilde Q} m_{\tilde U}$. The loop functions $f_1$ and $f_2$ depend on the mass ratios $x_2 = M_2^2/\tilde m^2$ and $x_\mu = \mu^2/\tilde m^2$ and their analytical form is given in the appendix.
Concerning the structure of eq.~(\ref{eq:CLnu}), we note that among the required two $SU(2)_L$ breaking insertions in the $Z$ penguin, one is formally provided by the helicity and flavour changing mass insertion $(\delta_u^{RL})_{32}$ and the other one by a Higgsino-Wino mixing (diagram $a$) or a flavour conserving helicity flip for the stop (diagram $b$), respectively.

To summarize, the contributions to $C_R^\nu$ in the MSSM turn out to be very small which implies that $\eta \simeq 0$ and that the longitudinal polarization fraction in the $B \to K^* \nu\bar\nu$ decay, $F_L(s_B)$, is always SM like. However, visible effects in $C_L^\nu$ can still be generated by chargino contributions through a large $(\delta_u^{RL})_{32}$ mass insertion.
For the numerical analysis we therefore choose an MSSM scenario where exactly such chargino effects are pronounced.
In particular, as these chargino contributions are not sensitive to the value of $\tan\beta$, we choose to work in the low $\tan\beta$ regime, thereby avoiding possible large Higgs effects in $B_s \to \mu^+ \mu^-$ and the corresponding constraint from this decay. We scan the relevant MSSM parameters in the following ranges
\begin{eqnarray}
5 < \tan\beta < 10~~&,&~~~ m_{\tilde Q}, m_{\tilde U}, M_2 < 1\textnormal{TeV}~, \nonumber \\
-1\textnormal{TeV} < \mu < 1\textnormal{TeV}~~&,&~~~ -3 < A_t/\sqrt{m_{\tilde Q} m_{\tilde U}} < 3~, \nonumber \\
0 < |(\delta_u^{RL})_{32}| < 1~~&,&~~~ 0 < \textnormal{Arg}\left[(\delta_u^{RL})_{32}\right] < 2\pi
\end{eqnarray}
and fix the remaining mass parameters to 1~TeV. We apply the existing constraints coming from direct searches for SUSY particles, from the lower bound on the Higgs mass, from the absence of charge and color breaking minima in the scalar potential as well as from the measurements of various FCNC processes like $B \to X_s\gamma$, $B \to X_s \ell^+ \ell^-$, $\Delta M_s / \Delta M_d$, $\epsilon_K$ and $\Delta M_K$.

\begin{figure}[tb]
\centering
\includegraphics[width=\textwidth]{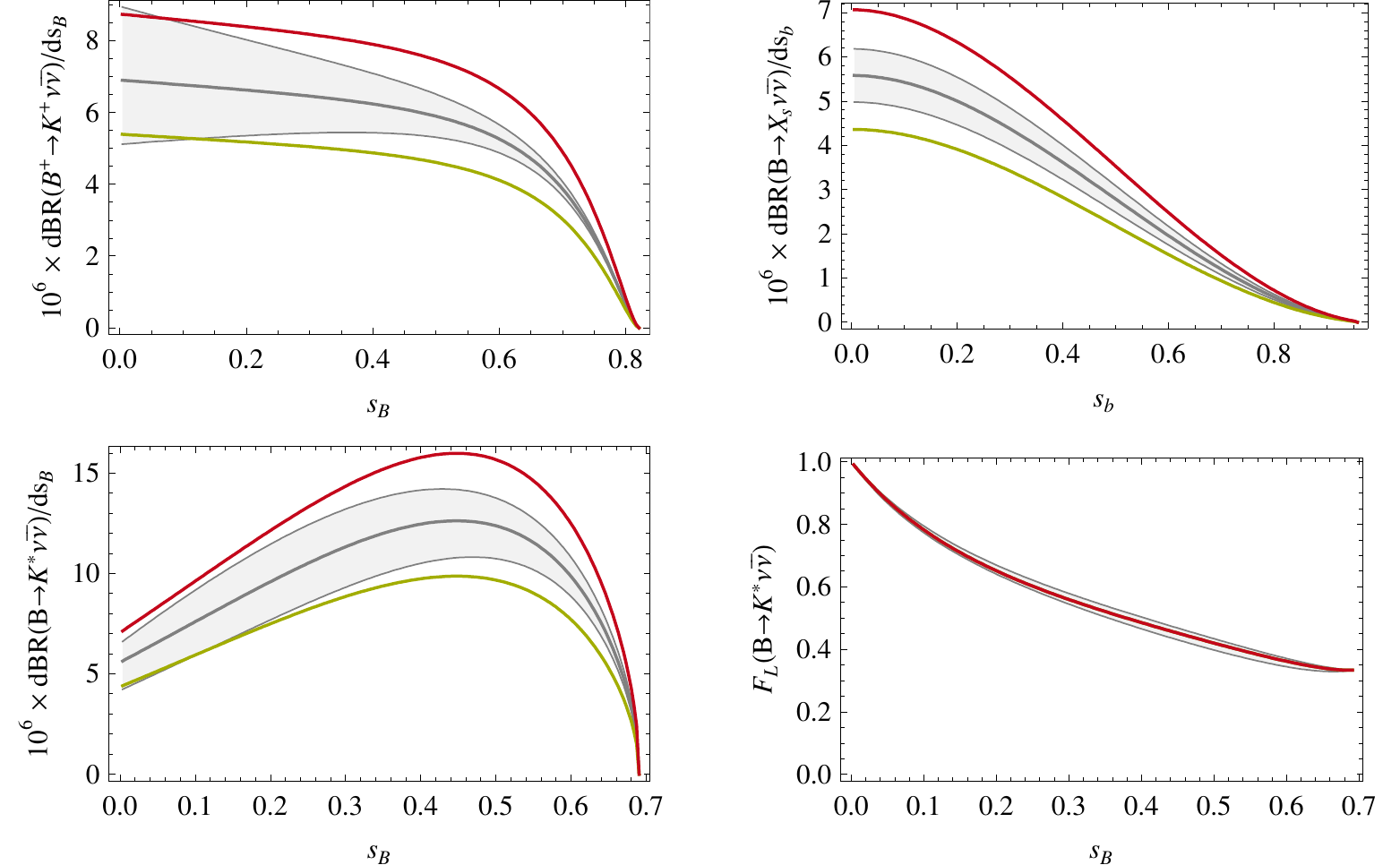}
\caption[]{\small Dependence of the four $b\to s\nu\bar\nu$ observables on the normalized neutrino invariant masses squared $s_{b,B}$ for two MSSM parameter points that give large effects within the considered scenario. The upper red lines correspond to the MSSM parameter set I of table~\ref{tab:MSSMparameter}, while the lower green ones correspond to parameter set II. The gray bands represent the SM predictions and the corresponding theory uncertainty.}
\label{fig:MSSM}
\end{figure}

Within that setup we obtain the following ranges for the branching ratios of the decays $B \to K^* \nu\bar\nu$, $B \to K \nu\bar\nu$ and $B \to X_s \nu\bar\nu$
\begin{eqnarray}
\label{eq:BKsnn_MSSM}
5.3 \times 10^{-6} \lesssim ~&~ \text{BR}(B \to K^* \nu\bar\nu) ~&~ \lesssim 8.7 \times 10^{-6}~, \\
\label{eq:BKnn_MSSM}
3.5 \times 10^{-6} \lesssim ~&~ \text{BR}(B \to K \nu\bar\nu) ~&~ \lesssim 5.8 \times 10^{-6}~, \\
\label{eq:BXsnn_MSSM}
2.1 \times 10^{-5} \lesssim ~&~ \text{BR}(B \to X_s \nu\bar\nu) ~&~ \lesssim 3.6 \times 10^{-5}~,
\end{eqnarray}
and we stress that due to the absence of significant effects in $C_R^\nu$ these three branching ratios are perfectly correlated. The effects in the corresponding differential branching ratios for these decays are shown in figure~\ref{fig:MSSM} for the two example MSSM parameter sets given in table~\ref{tab:MSSMparameter}.

\begin{table}[tb]
\centering
\renewcommand{\arraystretch}{1.5}
\begin{tabular}{|l|c|c|c|c|c|c|c|}
\hline
Parameter Set  & $\tan\beta$  & $\mu$ & $M_2$ & $m_{\tilde Q}$ & $m_{\tilde U}$ & $A_t$  & $(\delta_u^{RL})_{32}$ \\
\hline\hline
I              & 5            & $500$ & $800$ & $500$          & $400$          & $-800$ & $0.75$ \\
\hline
II             & 5            & $120$ & $700$ & $400$          & $800$          & $-700$ & $-0.5$ \\
\hline
\end{tabular}
\renewcommand{\arraystretch}{1}
\caption[]{\small Two example MSSM parameter sets giving large effects in $b\to s\nu\bar\nu$ transitions. Dimensionful quantities are expressed in GeV.}
\label{tab:MSSMparameter}
\end{table}

\subsubsection[Correlation with $B_s \to \mu^+ \mu^-$]{\boldmath{Correlation with $B_s \to \mu^+ \mu^-$}}

\begin{figure}[tb]
\centering
\includegraphics[width=0.5\textwidth]{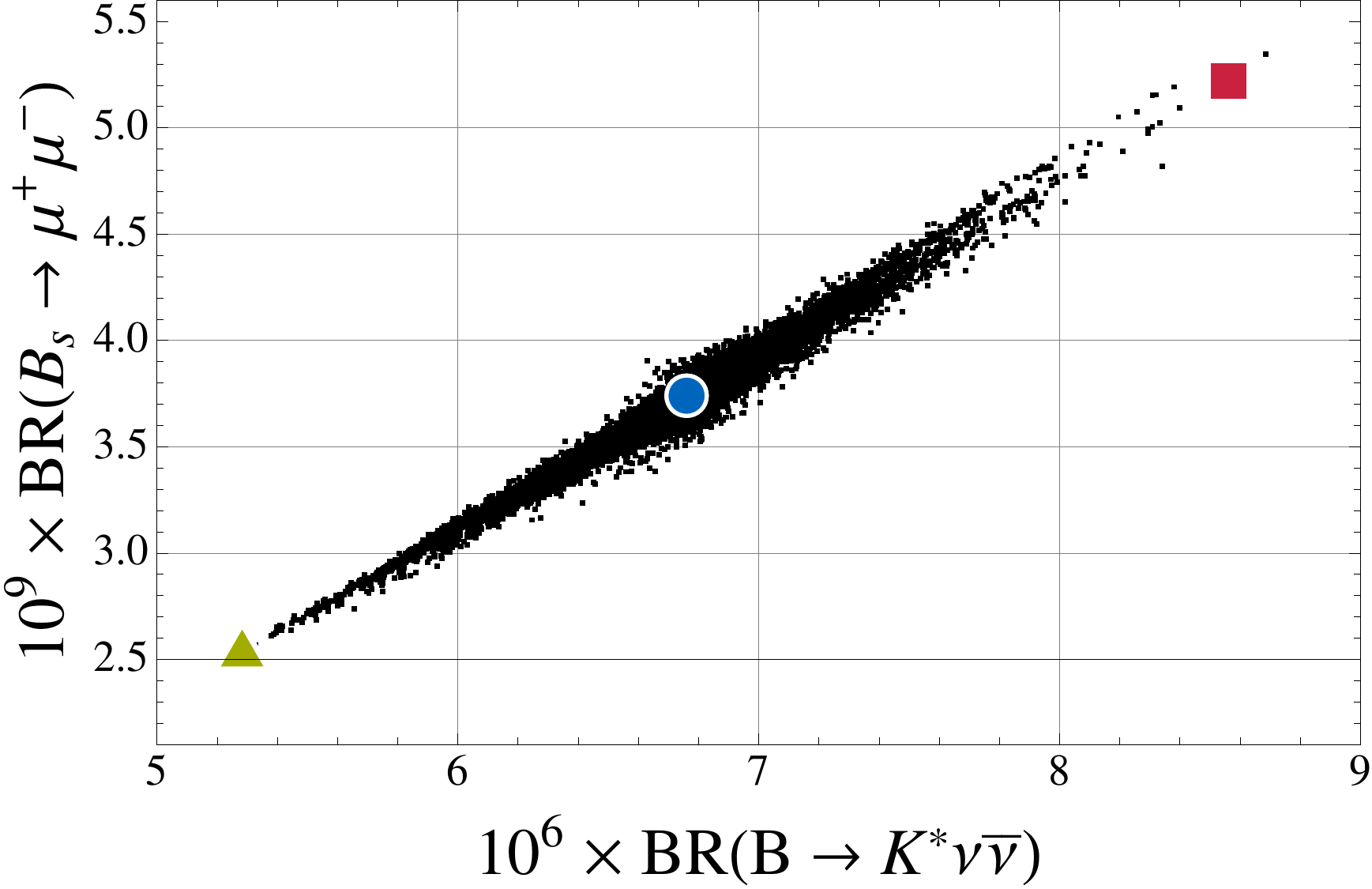}
\caption[]{\small Correlation between BR$(B \to K^* \nu\bar\nu)$ and BR$(B_s \to \mu^+ \mu^-)$ in the considered MSSM scenario. The blue circle represents the SM point, while the red square (green diamond) corresponds to the MSSM parameter set I (II). }
\label{fig:Bsmumu}
\end{figure}

In figure~\ref{fig:Bsmumu} we show the correlation between the branching ratios of $B \to K^* \nu\bar\nu$ and $B_s \to \mu^+ \mu^-$. This correlation arises because of the dominant contributions of $Z$ penguins to these two processes. We stress here that in our framework $\tan\beta$ is small and the heavy Higgs masses are fixed to 1~TeV, as slepton masses are. This leads both to negligible Higgs penguin and box contributions to $B_s \to \mu^+\mu^-$. A deviation from the shown correlation would thus point either towards sizable box contributions to $B \to K^* \nu\bar\nu$ or $B_s \to \mu^+ \mu^-$, which is possible with a very light slepton spectrum, or towards the presence of Higgs penguins in the $B_s \to \mu^+ \mu^-$ decay.

\subsubsection[Correlation with $K_L\to \pi^0\nu\bar\nu$ and $K^+\to \pi^+\nu\bar\nu$]{\boldmath{Correlation with $K_L\to \pi^0\nu\bar\nu$ and $K^+\to \pi^+\nu\bar\nu$}}

Within the chosen framework, we also investigate correlations between the decays $B \to K^* \nu\bar\nu$, $B \to K \nu\bar\nu$ and $B \to X_s \nu\bar\nu$ on the one side and $K_L\to \pi^0\nu\bar\nu$ and $K^+\to \pi^+\nu\bar\nu$ on the other side, that is correlations between $b \to s\nu\bar\nu$ and $s \to d\nu\bar\nu$ transitions \cite{Buras:2001af}. As we only switch on a mass insertion that corresponds to a $b\to s$ flip one might expect that effects in $K_L\to \pi^0\nu\bar\nu$ and $K^+\to \pi^+\nu\bar\nu$ are quite limited. However, as it turns out, also the considered $(\delta_u^{RL})_{32}$ mass insertion alone can induce large effects in the Kaon decays \cite{Isidori:2006qy}.

Before analysing these effects in more detail, we first summarize the theoretical description of the decays $K_L\to \pi^0\nu\bar\nu$ and $K^+\to \pi^+\nu\bar\nu$, for details see e.g. \cite{Buras:2004uu}. The effective Hamiltonian relevant for these decays in the context of the MSSM reads
\begin{equation}\label{hMSSM} 
{\mathcal H}_{\rm eff}=- \frac{4 G_{\rm F}}{\sqrt 2} \left[ 
{\mathcal H}_{\rm eff}^{(c)}+V^{\ast}_{ts}V_{td} \left(C^K_L \mathcal{O}^K_L + C^K_R \mathcal{O}^K_R\right)\right]
 ~+~ {\rm h.c.}~,
\end{equation}
where ${\mathcal H}_{\rm eff}^{(c)}$ denotes the operators which encode physics below the electroweak scale and the other term denotes the part of the  effective Hamiltonian sensitive to short-distance dynamics. The operators in eq.~(\ref{hMSSM}) read
\begin{align}
 \mathcal{O}^K_L=\frac{e^2}{16\pi^2}(\bar{s} \gamma_{\mu} P_L b)( \bar{\nu} \gamma^\mu(1- \gamma_5) \nu)~,~~&
 \mathcal{O}^K_R=\frac{e^2}{16\pi^2}(\bar{s} \gamma_{\mu} P_R d)( \bar{\nu} \gamma^\mu(1- \gamma_5) \nu)~.
\end{align}
The branching ratios can then be written as follows
\begin{eqnarray}
\label{bkpnZ}
{\rm BR}(K^+\to \pi^+\nu\bar\nu) ~~&=&~~ \kappa_+\left[\left(\frac{{\rm Im}(\lambda_t X^K)}{\lambda^5}\right)^2 
+ \left( - P_{(u,c)} + \frac{{\rm Re}(\lambda_t X^K)}{\lambda^5}\right)^2\right]~, \\
\label{bklpnZ}
{\rm BR}(K_L\to \pi^0\nu\bar\nu ) ~~&=&~~ \kappa_L \left( \frac{{\rm Im}(\lambda_t X^K)}{\lambda^5}\right)^2~,
\end{eqnarray}
where $X^K$ denotes the combination $-s_w^2(C^K_L + C^K_R)$. For the $\kappa$-factors, which originate mainly from hadronic matrix elements, we use $\kappa_+ = (5.27\pm 0.03) \times  10^{-11}$ and $\kappa_L = (2.27\pm 0.01) \times 10^{-10}$ \cite{Buras:2004uu,Mescia:2007kn}. Furthermore, we take $P_{(u,c)} = 0.41\pm 0.05$ \cite{Buras:2004uu,Isidori:2005xm,Buras:2005gr,Buras:2006gb,Brod:2008ss}, which accounts for contributions from charm and light quark loops. 

The $K_L\to \pi^0\nu\bar\nu$ and $K^+\to \pi^+\nu\bar\nu$ decays have been analysed in the MSSM by many authors \cite{Nir:1997tf,Buras:1997ij,Colangelo:1998pm,Buras:1999da,Buras:2004qb,Isidori:2006jh,Isidori:2006qy} and huge effects are still possible in particular coming from chargino contributions driven by a double $(\delta_u^{LR})_{13}(\delta_u^{RL})_{32}$ mass insertion \cite{Colangelo:1998pm} or from Higgs contributions in the large $\tan\beta$ regime \cite{Isidori:2006jh}.

\begin{figure}[tb]
\centering
\includegraphics[width=0.5\textwidth]{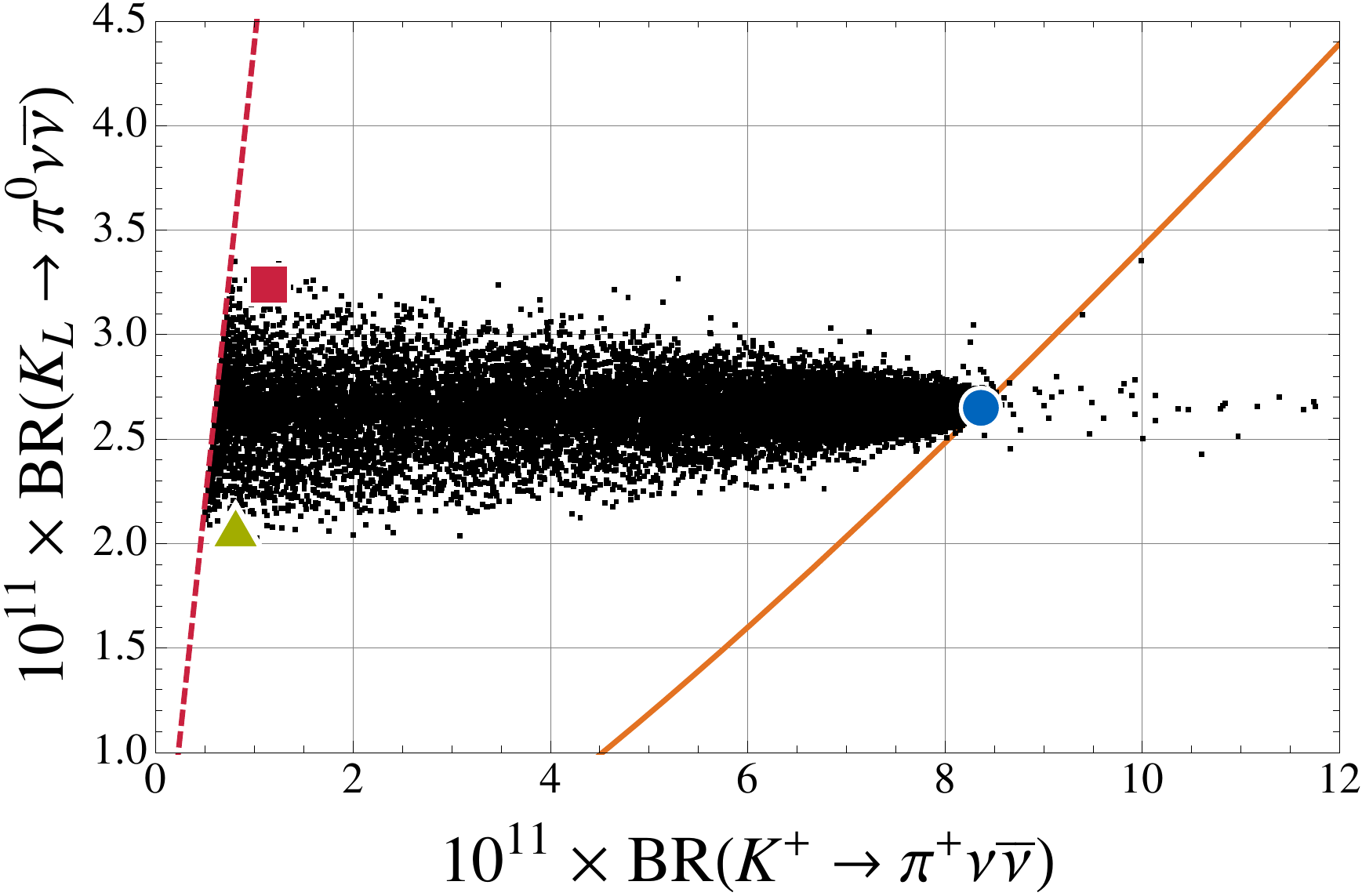}
\caption[]{\small BR$(K^+\to \pi^+\nu\bar\nu)$ vs. BR$(K_L\to \pi^0\nu\bar\nu)$. The blue circle shows the SM prediction and the red square (green triangle) corresponds to the MSSM parameter set I (II). The dashed red line represents the Grossman-Nir bound \cite{Grossman:1997sk}, while the solid orange line shows the correlation in models with MFV.}
\label{fig:Kdecays}
\end{figure}

\begin{figure}[tb]
\centering
\includegraphics[width=0.99\textwidth]{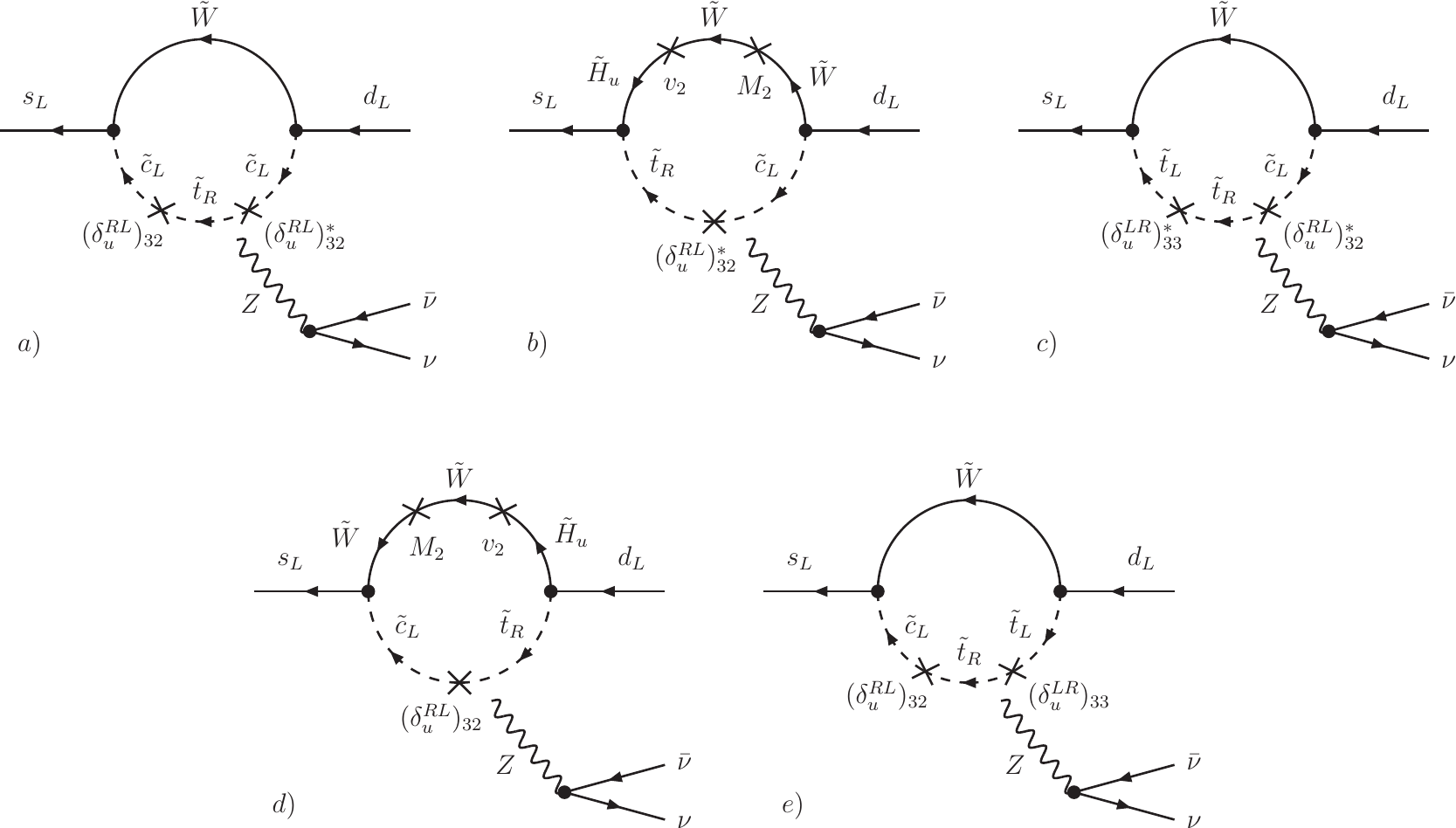}
\caption[]{\small Dominant chargino contributions to the Wilson coefficient $C_L^K$ in the mass insertion approximation.}
\label{fig:Kdiagrams}
\end{figure}

The effects in $K_L\to \pi^0\nu\bar\nu$ and $K^+\to \pi^+\nu\bar\nu$ we find in the considered MSSM framework are shown in figure~\ref{fig:Kdecays}. In particular, while the branching ratio of $K_L\to \pi^0\nu\bar\nu$ is changed by at most $\simeq \pm 20\%$, the branching ratio of $K^+\to \pi^+\nu\bar\nu$ turns out to be significantly reduced and can reach values as low as allowed by the model independent Grossman-Nir bound \cite{Grossman:1997sk}.
The dominant chargino contributions to the Wilson coefficient $C_L^K$ that are responsible for these effects are shown in figure~\ref{fig:Kdiagrams} and they are given by the following approximate expression

\begin{eqnarray}
\label{eq:CLs}
C_L^K &~~\simeq~~& \frac{1}{s_w^2} \frac{V_{cd}V_{cs}^*}{V_{td}V_{ts}^*}  |(\delta_u^{RL})_{32}|^2 \frac{1}{8} f_1(x_2) \nonumber \\
      &   ~~~~   & - \frac{1}{s_w^2} \left( \frac{V_{cs}^*}{V_{ts}^*} (\delta_u^{RL})_{32} + \frac{V_{cd}}{V_{td}} (\delta_u^{RL})^*_{32}\right) \left[ \frac{m_t A_t}{8 \tilde m^2} f_1(x_2) - \frac{m_t M_2}{4 \tilde m^2} f_2(x_\mu,x_2) \right]~.
\end{eqnarray}
As we only consider a $b\to s$ mass insertion, contributions to the Kaon decays necessarily involve also additional CKM matrix elements. The leading contribution receives a CKM suppression only from the matrix element $V_{cd}$ and it is shown in diagram a) of figure~\ref{fig:Kdiagrams}. It gives an almost real contribution to the combination $\lambda_t C_L^K$ that is proportional to $V_{cd}V_{cs}^*|(\delta_u^{RL})_{32}|^2$ and therefore dominantly leads to effects in BR$(K^+\to \pi^+\nu\bar\nu)$. 
Diagrams b) to e) of figure~\ref{fig:Kdiagrams} on the other hand involve also $V_{ts}$ and $V_{td}$ and introduce sensitivity to the phase of $(\delta_u^{RL})_{32}$ and the phase of $V_{td}$ and can therefore also affect BR$(K_L\to \pi^0\nu\bar\nu)$. However these contributions are suppressed compared to the one of diagram a) roughly by a factor $\lambda^2$ and therefore the effects in BR$(K_L\to \pi^0\nu\bar\nu)$ are generically smaller than in BR$(K^+\to \pi^+\nu\bar\nu)$.

We find that the $(\delta_u^{RL})_{32}$ mass insertion can also lead to sizable effects in the $K^0 - \bar K^0$ mixing amplitude. The leading contribution is proportional to $|(\delta_u^{RL})_{32}|^4 (V_{cs} V_{cd}^*)^2$ and can become comparable to the dominant SM charm contribution. In our numerical analysis we included therefore also the constraints from $\Delta M_K$ and $\epsilon_K$.

\begin{figure}[tb]
\centering
\includegraphics[width=\textwidth]{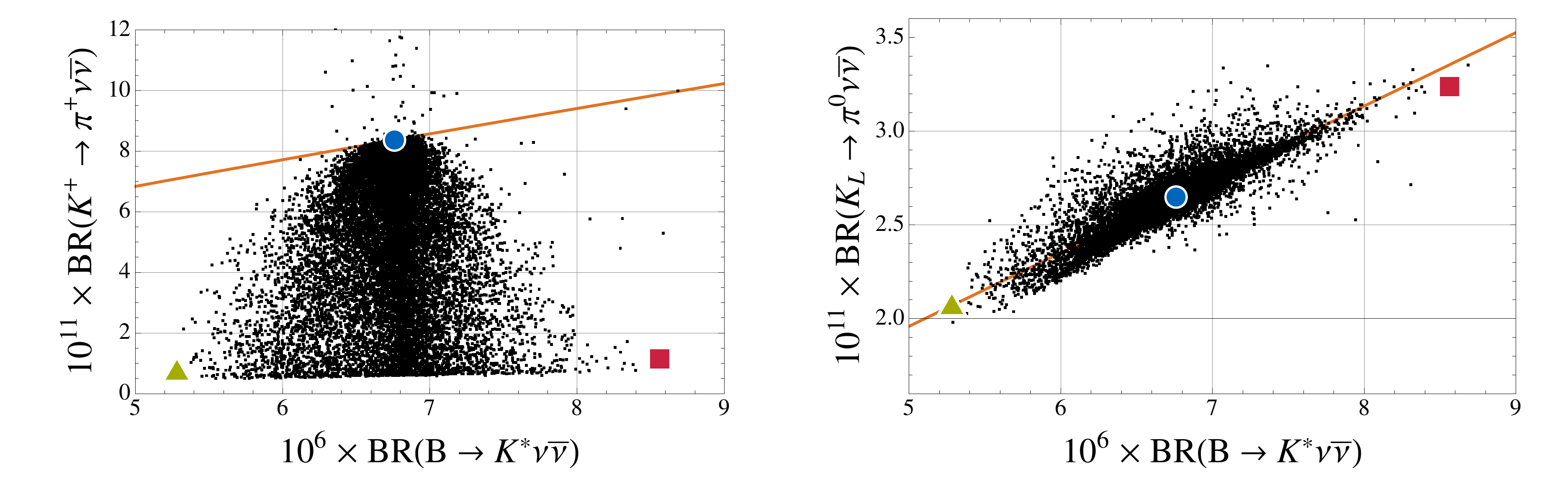}
\caption[]{\small Left plot: BR$(B \to K^* \nu\bar\nu)$ vs. BR$(K^+\to \pi^+\nu\bar\nu)$. Right plot: BR$(B \to K^* \nu\bar\nu)$ vs. BR$(K_L\to \pi^0\nu\bar\nu)$. The blue circles represent the SM predictions, while the red squares (green diamonds) correspond to the MSSM parameter set I (II). The solid orange lines show the correlations in models with MFV.}
\label{fig:BvsKdecays}
\end{figure}

As both the effects in $b\to s \nu\bar\nu $ and $s\to d \nu\bar\nu$ are generated by the same mass insertion $(\delta_u^{RL})_{32}$ one in fact expects correlations between the $B$ and $K$ decays. However, we remark that the corresponding branching ratios show a different behavior with respect to the mass insertion $(\delta_u^{RL})_{32}$. The $B$ decays are mostly sensitive to the real part of the mass insertion, while $K^+\to \pi^+\nu\bar\nu$ receives the dominant contribution from its absolute value. The branching ratio of $K_L\to \pi^0\nu\bar\nu$ finally depends mainly on the imaginary part of the combination $V_{ts}^*V_{cd}(\delta_u^{RL})_{32}^* + V_{cs}^*V_{td} (\delta_u^{RL})_{32}$. In figure~\ref{fig:BvsKdecays} we show the emerging correlations between the branching ratios of $B \to K^* \nu\bar\nu$ and of $K^+\to \pi^+\nu\bar\nu$ and $K_L\to \pi^0\nu\bar\nu$ and compare them to the corresponding correlations expected in models with MFV.

\subsection{Decay to invisible scalars}\label{sec:scalars}

We will now turn to analyse the impact of a simple extension of the low energy particle content of the SM. More precisely, we add an additional gauge-singlet scalar $S$ with mass $m_S<m_b/2$. The scalar could then be produced in $b\to s$ transitions and, assuming it to be stable or sufficiently long-lived to escape the detector, it would contribute to the $b\to s\nu\bar\nu$ observables, since the two final states could not be distinguished experimentally.
This setup finds application in models of dark matter \cite{Bird:2004ts}, but we emphasize that we do not assume any particular model generating the $b\to sSS$ transition.

\subsubsection{Effective theory}
The effective Hamiltonian describing the flavour-changing quark-scalar interaction can be written as
\begin{equation}
\mathcal H_\text{eff}  = C_L^S \frac{{m_b }}{2}(\bar sP_L b)S^2  + C_R^S \frac{{m_b }}{2}(\bar sP_R b)S^2~.
\end{equation}
In addition to the two Wilson coefficients $C_L^S$ and $C_R^S$, the mass of the scalar particle $m_S$ enters the observables through the phase space integration. We will consider $C_L^S$, $C_R^S$ and $m_S$ as independent parameters, although there are certain relations in specific high energy models, e.g. in the model of ref. \cite{Bird:2004ts}, where the Wilson coefficient $C_R^S$ is generated by a Higgs penguin.

\subsubsection{Corrections to the observables}\label{sec:scalar-obs}
The extra scalar final state leads to an additional contribution to the differential decay width of the three processes considered in our paper, since the $s\nu\bar\nu$ and $sSS$ final states cannot be distinguished experimentally. Therefore, one has
\begin{equation}
\frac{d\Gamma(B\to X_s \slashed E) }{ds_b } = \frac{d\Gamma(B\to X_s \nu\bar\nu) }{ds_b } + \frac{d\Gamma(B\to X_s SS) }{ds_b }~,
\end{equation}
and correspondingly for the exclusive decays. The differential decay widths of the scalar modes read
\begin{equation}
\frac{{d^2\Gamma(B\to K^*(\to K\pi)SS) }}{{ds_B d\cos\theta}} = m_B^5 \frac{3 A_0^2 (s_B )\left| {C_L^S  - C_R^S} \right|^2}{2^{12} \pi ^3 } \sqrt {1 - \dfrac{{4 \widetilde{m}_S^2 }}{{s_B }}} \lambda ^{3/2} (1 ,\widetilde{m}_{K^ *  }^2 ,s_B) \cos^2\theta~,
\label{eq:BKsSS}
\end{equation}
\begin{equation}
\frac{{d\Gamma(B\to KSS) }}{{ds_B}} = m_B^5 \frac{{\left[f_0^K (s_B)\right]^2\left( {1  - \widetilde{m}_K^2 } \right)^2 \left| {C_L^S  + C_R^S} \right|^2 }}{{2^{11} \pi ^3 }}\sqrt {1 - \dfrac{{4\widetilde{m}_S^2 }}{{s_B}}} \lambda ^{1/2} (1,\widetilde{m}_K^2 ,s_B)~,
\end{equation}
\begin{multline}
\frac{d\Gamma(B\to X_sSS)}{ds_b} = m_b^5\frac{\left| {C_R^S} \right|^2  + \left| {C_L^S } \right|^2}{2^{9} \pi ^3 }
 \sqrt {1 - \dfrac{{4\hat m_S^2 }}{{s_b}}} \lambda ^{1/2} (1 ,\hat m_s^2 ,s_b ) \\
\times \left[
\left( 1  + \hat m_s^2  - s_b  \right)
- 4 \hat m_s \frac{\text{Re}\left(C^S_L C_R^{S *}\right)}{\left| {C_R^S} \right|^2  + \left| {C_L^S } \right|^2} \right]
~,
\end{multline}
where $A_0(s_B)$ and $f_0^K(s_B)$ are the scalar $B\to K^*$ and $B\to K$ form factors\footnote{%
Note that, by abuse of notation, we use the symbol $A_0(s_B)$ for the scalar form factor in this section, while it was used for the longitudinal transversity amplitude in section~\ref{sec:BKsnunu}.
}, respectively. We obtain $A_0$ by the procedure described in section \ref{sec:BKsnunu}, while $f_0^K$ is taken from \cite{Ball:2004ye}.

The observable $F_L$, as it is extracted from the angular distribution of $B\to K^* (\to K\pi)\slashed E$ according to the formula (cf. eq.~(\ref{eq:doublediff}))
\begin{equation}
\frac{d^2\Gamma}{ds_B d\!\cos\!\theta}\bigg/\frac{d\Gamma}{ds_B} =  \frac{3}{4} (1-F_L) \sin^2\theta + \frac{3}{2} F_L \cos^2\theta~,
\end{equation}
is modified according to 
\begin{equation}
F_L(B\to K^* \slashed E) = \frac{ d\Gamma_L(B\to K^* \nu\bar\nu)/s_B + d\Gamma(B\to K^* SS)/s_B }{ d\Gamma(B\to K^* \nu\bar\nu)/s_B  + d\Gamma(B\to K^* SS)/s_B}~,
\end{equation}
since the $K^*$ is always produced with longitudinal polarization in the $B\to K^* SS$ decay, which is also the reason for the factor of $\cos^2\theta$ in eq.~(\ref{eq:BKsSS}).

\begin{figure}[tb]
\centering
\includegraphics[width=\textwidth]{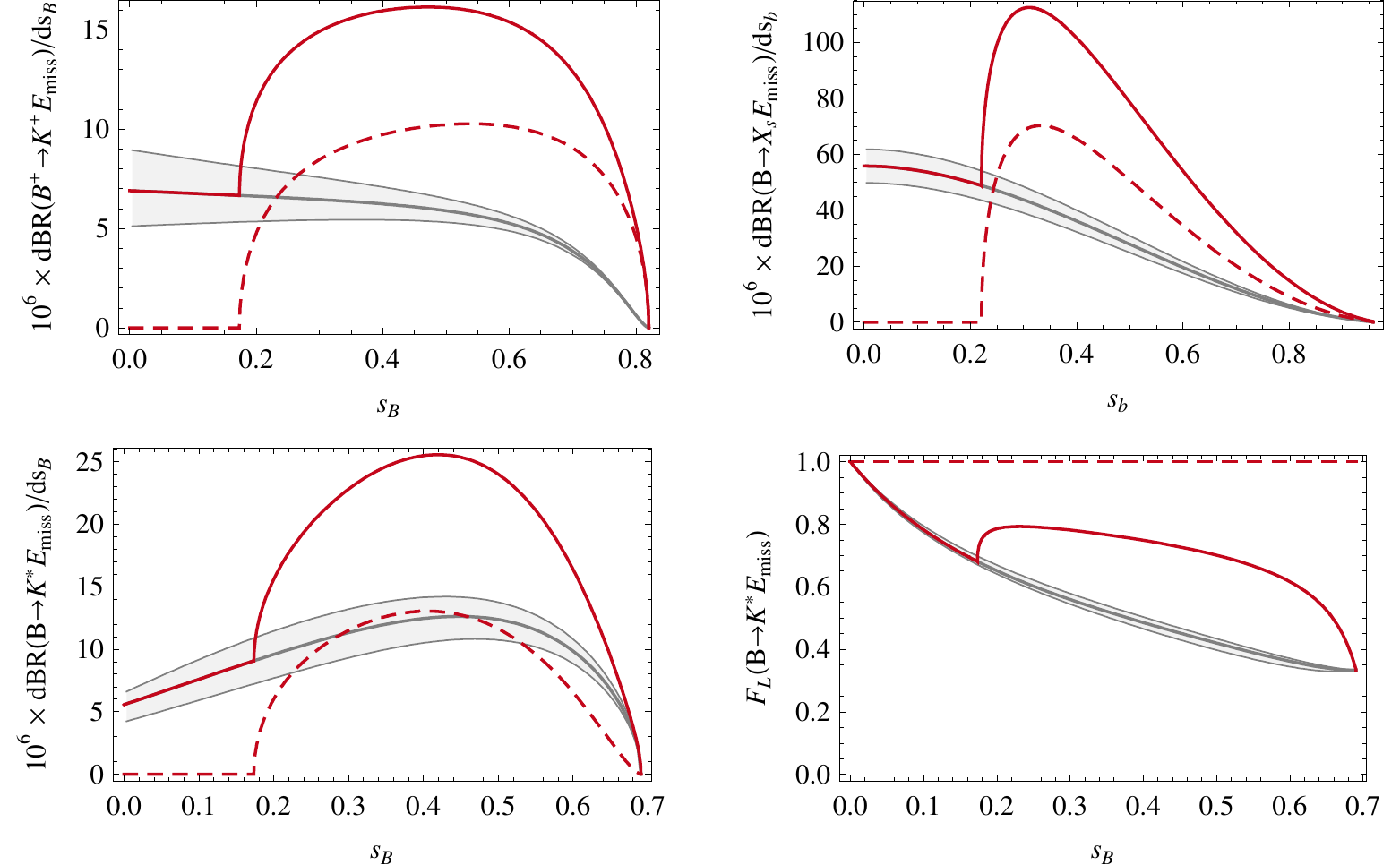}
\caption[]{\small Dependence of the four observables on the normalized neutrino invariant mass squared in a scenario in which SM-like $b\to s\nu\bar\nu$ processes overlap with $b\to sSS$ decays. The parameters chosen are $m_S=1.1$~GeV, $C_L^S=0$ and $C_R^S=2.8\times10^{-8}\;\text{GeV}^{-2}$. The grey curves show the pure $b\to s\nu\bar\nu$ (i.e. SM) contribution with theoretical uncertainties, the red dashed curves the pure $b\to sSS$ contribution and the red solid curves the resulting combination.
}
\label{fig:SS}
\end{figure}

\subsubsection{Numerical results}

The overlap of the decay distributions of $b\to s\nu\bar\nu$ and $b\to sSS$ decays leads to a characteristic spectrum with a kinematical edge at $q^2=m_S^2/4$ that would clearly signal the presence of an additional final state.
In figure~\ref{fig:SS}, we show the differential branching ratios of all three decays as well as $F_L(s_B)$ for a scenario in which $m_S=1.1$~GeV, $C_L^S=0$ and $C_R^S=2.8\times10^{-8}\;\text{GeV}^{-2}$ have been chosen such that all the branching ratios are well below their experimental upper bounds in table~\ref{tab:exp}.

\begin{figure}[tb]
\centering
\includegraphics[width=0.45\textwidth]{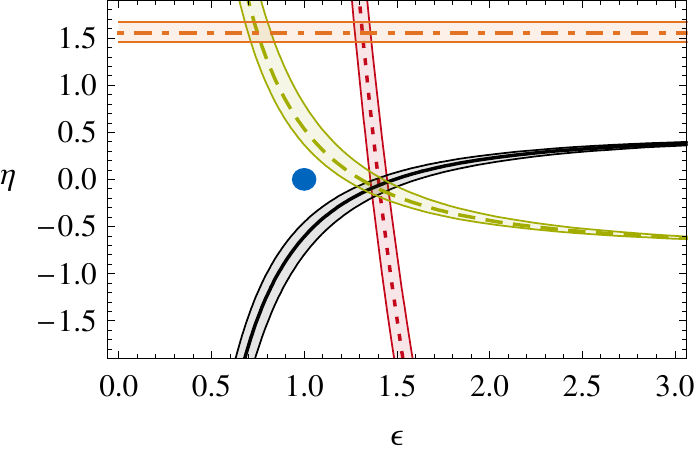}
\caption[]{\small Constraints on the $\epsilon$-$\eta$-plane obtained by applying eqs.~(\ref{eq:epseta-BKsnn})--(\ref{eq:epseta-FL}) in a scenario in which SM-like $b\to s\nu\bar\nu$ processes overlap with $b\to sSS$ decays. The parameters are chosen as in figure~\ref{fig:SS}. The colouring and dashing is as in figure \ref{fig:exp-hypo}.}
\label{fig:const-phantom}
\end{figure}

Due to the modification of the observables by the contributions in section \ref{sec:scalar-obs}, it is clear that eqs.~(\ref{eq:epseta-BKsnn})--(\ref{eq:epseta-FL}), relating the observables to the parameters $\epsilon$ and $\eta$, are no longer valid. In figure~\ref{fig:const-phantom}, we show the constraints on the $\epsilon$-$\eta$-plane which would result by naively applying eqs.~(\ref{eq:epseta-BKsnn})--(\ref{eq:epseta-FL}) anyway, with the parameter values chosen as above.
As a result, the bands corresponding to the different observables do not meet at a single point any longer. One observes that, while this splitting is quite small for the three branching ratios, the observable $\langle F_L \rangle$ displays unambiguously the invalidity of eqs.~(\ref{eq:epseta-BKsnn})--(\ref{eq:epseta-FL}). While, according to its definition in section~\ref{sec:indep}, $\eta$ is restricted to the interval $[-\frac{1}{2},\frac{1}{2}]$, its feigned value in this scenario, obtained by naively applying eq. (\ref{eq:epseta-FL}), can be bigger than $\frac{1}{2}$.

To summarize, in the presence of a light, stable (or long-lived) singlet scalar, the experimental measurements of the $b\to s\nu\bar\nu$ (or rather, $b\to s\slashed E$) observables can include contributions from invisible decays to scalars. Such decays would manifest themselves through characteristic kinematical edges in the spectra or through an inconsistency in the extraction of the parameters $\epsilon$ and $\eta$ from the different (integrated) observables. For this effect, which is reminiscent of the impact of a fourth generation of quarks on the unitarity triangle, the observable $\langle F_L \rangle$ turns out to be particularly useful.


\section{Summary}\label{sec:sum}

In this paper we have performed a new analysis of the decays $B\to K^*\nu\bar\nu$, $B\to K\nu\bar\nu$ and $B\to X_s\nu\bar\nu$ in the SM, model-independently and in a number of NP scenarios.

The novel features of our analysis are:
\begin{itemize}
\item Improved form factors entering $B\to K^*\nu\bar\nu$.
\item Improved estimate of the inclusive $\text{BR}(B\to X_s\nu\bar\nu)$ within the SM.
\item The introduction of the $(\epsilon,\eta)$ plane analogous to the $(\bar\varrho,\bar\eta)$ plane, known from CKM phenomenology, with a non-vanishing $\eta$ signalling this time not CP violation but the presence of right-handed down-quark flavour violating couplings.
\item Correlations between $b\to s\nu\bar\nu$ and $b\to s \ell^+\ell^-$ transitions.
\item Correlations between $b\to s\nu\bar\nu$ and $s\to d \nu\bar\nu$ transitions in non-MFV scenarios.
\end{itemize}

The three decays analysed here provide four global (integrated over the invariant mass $q^2$ of the $\nu\bar\nu$ pair) observables which can be chosen to be three branching ratios of the decays in question and one additional observable which can be obtained from $B\to K^*\nu\bar\nu$.

We have provided new SM predictions for these four global observables (table~\ref{tab:exp}) and the corresponding $q^2$ dependences (figure~\ref{fig:SM}).

Model-independently, the four observables can be expressed in terms of only two real parameters $\epsilon$ and $\eta$ so that measuring all four observables would overconstrain the resulting point in the $(\epsilon,\eta)$ plane with $(\epsilon,\eta)=(1,0)$ corresponding to the SM and $\eta\not=0$ signalling the presence of right-handed down-quark flavour violating couplings. As $\epsilon$ and $\eta$, being given directly in terms of the two Wilson coefficients $C^\nu_L$ and $C^\nu_R$, can be calculated straightforwardly in any NP scenario, the $(\epsilon,\eta)$ plane is very suitable for a transparent exhibition and comparison of various extensions of the SM. 

Performing an extensive numerical analysis of various NP scenarios we can provide the following messages:
\begin{itemize}
\item Our improved SM prediction $\text{BR}(B \to K^* \nu\bar\nu)=( 6.8^{+1.0}_{-1.1} ) \times 10^{-6}$ is significantly lower than the ones present in the literature.
\item Our calculation of $\text{BR}(B \to X_s \nu\bar\nu)=( 2.7 \pm 0.2 ) \times 10^{-5}$ in the SM is  considerably more accurate than the ones present in the literature.
\item Sizable deviations from the SM expectations are possible in the presence of significantly modified $Z$ penguins constrained mainly by the data on $b\to s \ell^+\ell^-$ transitions. Interesting correlations between various $b\to s\nu\bar\nu$ branching ratios and BR$(B\to X_s \ell^+\ell^-)$ follow (figure~\ref{fig:nunu-vs-ll}).
\item NP effects in the LHT model in which $\eta=0$ are found to be small. Also NP effects in the considered decays in a RS model with custodial protection of left-handed $Z$-couplings are small.
\item Sizable NP effects are found in the MSSM with a generic flavour violating soft sector constrained mainly by the data on $B \to X_s \gamma$ and $B_s\to \mu^+\mu^-$. The dominant NP contributions come from chargino effects in $C_L^\nu$. An interesting correlation between BR$(B\to K^*\nu\bar\nu)$ and BR$(B_s\to \mu^+\mu^-)$ (figure~\ref{fig:Bsmumu}) offers a useful test of a particular MSSM scenario in this class of supersymmetric models.
\item The known strong correlation between $b\to s\nu\bar\nu$ and $s\to d \nu\bar\nu$ transitions characteristic for CMFV models can be significantly violated in the MSSM with non-MFV interactions.
\item The impact of the presence of invisible scalars that could be produced in $b\to s$ transitions can be depicted transparently in the $(\epsilon,\eta)$ plane (figure~\ref{fig:const-phantom}) and implies a characteristic pattern of modifications in the $q^2$ distributions (figure~\ref{fig:SS}).
\end{itemize}

In summary, while our analysis of $b\to s \nu\bar\nu$ transitions does not allow to expect NP effects to be as spectacular as in $B\to K^*\ell^+ \ell^-$ analysed by us recently, the simultaneous analysis of the four basis observables that can be measured in these transitions, in particular in conjunction with the $(\epsilon,\eta)$ plane, offers useful means for tests of those NP physics scenarios in which $Z$-penguin contributions are significantly modified, non-MFV interactions are present and new right-handed down quark couplings are sizable.


\acknowledgments{
The authors would like to thank
Miko{\l}aj Misiak and Martin Gorbahn for illuminating comments on the estimation of the $B\to X_s\nu\bar\nu$ theory uncertainty,
Patricia Ball and Aoife Bharucha for useful discussions in the initial stages of this work,
Thorsten Feldmann and Paride Paradisi for many helpful discussions,
as well as Andr\'e Hoang and Gudrun Hiller for comments on the final manuscript.
This work has been supported in part by the Cluster of Excellence ``Origin and Structure of the Universe'' and the German Bundesministerium f\"ur Bildung und Forschung under contract 05HT6WOA.
}


\begin{appendix}
\section{Loop functions}
Here we give the analytical expressions for the loop functions that appear in the Wilson coefficients of section~\ref{NPMSSM}.

\begin{eqnarray}
f_1(x) ~~&=&~~ \frac{1-5x-2x^2}{6(1-x)^3} - \frac{x^2}{(1-x)^4} \log(x)~,
\end{eqnarray}
\begin{eqnarray}
f_2(x,y) ~~&=&~~ - \frac{x(2x^2-x(y+3)+2y)}{2(1-x)^3(y-x)^2}\log(x) - \frac{2x^2(y+1)-5xy^2+y^2(2y-1)}{2(1-y)^3(x-y)^2}\log(y) \nonumber \\
~~&&~~ + \frac{8x^3-x^2(7y+11)+x(y(y+10)+1)-y(3y-1)}{4(1-x)^2(1-y)^2(y-x)}~.
\end{eqnarray}

\end{appendix}


\bibliography{absw}
\bibliographystyle{JHEP}

\end{document}